\documentclass[a4paper,11pt]{article}
\pdfoutput=1 

\usepackage{jcappub} 

\usepackage[T1]{fontenc} 

\title{\boldmath Thermal-acoustic fermionic dynamics and primordial perturbations in warm Yukawa inflation}

\author[a,b,1]{Xi-Bin Li, \note{Corresponding author.}}
\author[a,b]{Ke Fu,}
\author[a,b,2]{Ya-Ting Wang, \note{Corresponding author.}}
\author[a,b]{De-Tian Zhao, }
\author[c,3]{and Xiao-Min Zhang \note{Corresponding author.}}

\affiliation[a]{College of Physics and Electronic Information, Inner Mongolia Normal University, 81 Zhaowuda Road, Hohhot, 010022, Inner Mongolia, China}
\affiliation[b]{Inner Mongolia Key Laboratory of Applied Condensed Matter Physics, Inner Mongolia Normal University, 81 Zhaowuda Road, Hohhot, 010022, Inner Mongolia, China}
\affiliation[c]{School of Science, Qingdao University of Technology, Qingdao 266033, China}

\emailAdd{lxbimnu@imnu.edu.cn}
\emailAdd{1617533168@qq.com}
\emailAdd{ytwang@imnu.edu.cn}
\emailAdd{zhaodetian1003@163.com}
\emailAdd{zhangxm@mail.bnu.edu.cn}

\abstract{We investigate fermion production and its cosmological consequences in a warm inflationary scenario with a Yukawa interaction between the inflaton field and Dirac fermions. 
Treating the produced fermions as a thermalized relativistic fluid, we derive the coupled perturbation equations incorporating dissipative effects and thermal fluctuations. 
By solving the resulting system through Green’s function methods, we identify distinct contributions from thermal noise, thermal–acoustic propagation, and higher-order scattering processes to the scalar perturbations. 
We show that acoustic propagation of fermionic fluctuations becomes significant in the relativistic regime, 
where the competition between gravitational attraction and Pauli degeneracy pressure allows collective fermionic modes to influence the primordial spectrum. 
Furthermore, we calculate the fermion number density and acoustic velocity, revealing that strong Yukawa interactions enhance non-adiabatic fermion production and modify the thermodynamic properties of the fermion bath. 
The sourced scalar and tensor perturbations are calculated, indicating that fermion production can substantially amplify primordial fluctuations, particularly in the strong-coupling regime. 
Since scalar perturbations receive more efficient enhancement than tensor modes, the tensor-to-scalar ratio is suppressed relative to the standard single-field warm inflation prediction. Our results show that thermalized fermions provide an additional mechanism for modifying primordial perturbations and offer new insights into the role of particle production in warm inflationary cosmology.}

\begin{document}
\maketitle
\flushbottom

\section{\label{introduction}Introduction}

Cosmic inflation provides a successful framework for describing the accelerated expansion of the early Universe and the generation of primordial fluctuations that subsequently seed large-scale structure. 
Precision measurements of the cosmic microwave background (CMB) have increasingly constrained the scalar spectral index and the amplitude of primordial tensor perturbations, 
making the tensor-to-scalar ratio $r$ an important probe of the inflationary energy scale and of interactions active during inflation. 
Recent theoretical and numerical developments have substantially improved the treatment of cosmological perturbations in warm inflation \cite{warm1,warm2,warm3,warm_insert1,warm_insert2,warm_insert3}, 
while increasingly precise cosmological observations provide new opportunities to distinguish dissipative inflationary scenarios from their cold-inflation counterparts.

In contrast to the conventional cold-inflation picture, the inflaton continuously dissipates energy into other degrees of freedom during the accelerated stage, sustaining a non-negligible thermal bath. 
Dissipation modifies the background evolution of the inflaton, while the associated fluctuation-dissipation relation introduces stochastic thermal fluctuations that can contribute directly to primordial perturbations \cite{warm_insert4,warm_insert5}. 
During the past few years, warm inflation has been investigated in a wide range of settings, including noncanonical dynamics \cite{warm4}, ultraslow-roll and constant-roll evolution \cite{warm5,warm6,warm7}, 
Higgs inflation \cite{warm8}, chromonatural inflation \cite{warm9}, multi-natural inflation \cite{warm10, warm11}, and string-motivated constructions \cite{warm12}. 
These studies demonstrate that thermal and dissipative effects are not merely phenomenological corrections but can qualitatively modify the inflationary dynamics and its observable predictions.

Considerable attention has also been devoted to establishing the microscopic particle-physics origin of warm inflation \cite{warm13,warm14}. 
Dissipation may arise from explicit interactions between the inflaton and scalar, fermionic, or gauge degrees of freedom. 
Recent developments have shown that viable high-temperature warm-inflation models can be constructed with particle contents. 
In particular, the Warm Little Inflaton framework can be realized through dissipative interactions involving fermionic degrees of freedom \cite{warm5,warm15}, 
while more recent studies have explored warm inflation based on Standard Model interactions and have examined its viability with precision numerical calculations \cite{warm16,warm17,warm18}. 
These developments strengthen the motivation for studying explicitly how microscopic particles produced by the inflaton participate not only in the background thermal bath but also in cosmological perturbations.

Fermionic degrees of freedom are particularly interesting because the inflaton can couple directly to a Dirac field through the Yukawa interaction $-g\phi\bar{\psi}\psi$, 
which generates a field-dependent effective fermion mass, $ m_{\rm eff}=m_\psi+g\phi$. 
Fermion production in an expanding spacetime has a long theoretical history, but its inflationary phenomenology continues to receive considerable attention \cite{Yukawa1,Yukawa2,Yukawa3,Yukawa4}. 
A recent analysis of fermionic inflation with a direct Yukawa interaction showed that the effective mass and Yukawa coupling determine the non-adiabaticity of fermion production and can substantially modify the resulting tensor-to-scalar ratio \cite{Dirac4,Yukawa5}. 
More generally, recent work on quark, lepton, and right-handed-neutrino production during inflation has further emphasized the sensitivity of fermion production to masses generated through Yukawa couplings \cite{warm17,Yukawa5,warm19,Yukawa6,Yukawa7,warm_insert6}.

This issue becomes particularly relevant in warm inflation, where the produced particles can interact with an already existing thermal environment. 
The question is then no longer limited to the number of particles produced from the inflationary background and recent studies have increasingly emphasized the importance of thermalization during inflation.
Nevertheless, existing investigations have largely focused on the microscopic origin of dissipation, the background dynamics of warm inflation, particle abundances, backreaction effects, or directly sourced perturbations. 
By contrast, the dynamical consequences arising from the produced particles have received comparatively little attention. 
In particular, the collective acoustic dynamics of thermalized fermions and their imprint on primordial perturbations remain comparatively unexplored. 
The intermediate physical process by which thermal fluctuations propagate acoustically through a Yukawa-produced fermion fluid before contributing to the curvature perturbation 
has not been systematically incorporated into the fermionic warm-inflation framework. This provides the central motivation for the present study.

In this work, we investigate thermal-acoustic fermionic dynamics in a warm inflationary model in which a scalar inflaton is directly coupled to Dirac fermions through a Yukawa interaction. The thermalized fermions are treated as a constituent of the thermal bath. Starting from total energy-momentum conservation, we derive the coupled perturbation equations including dissipative and stochastic thermal effects. We then reformulate the scalar perturbation problem as a Volterra integral equation and employ Green's functions to distinguish direct thermal propagation, thermal-acoustic propagation, and higher-order mixed processes. The resulting formulation explicitly connects the acoustic properties of the fermion fluid with the primordial curvature perturbation.

This paper is organized as follows. In Sec. \ref{spectra}, we introduce the warm Yukawa inflationary framework and derive the coupled dynamical equations for the inflaton and thermalized fermion fluid. 
In Sec. \ref{P_large}, we calculate the sourced scalar perturbations using the Green-function approach and investigate the thermal-acoustic propagation mechanism. 
Section \ref{sec_nf} is devoted to the fermion particle density and acoustic velocity and their dependence on the effective mass and Yukawa coupling. 
In Sec. \ref{sec_GW}, we evaluate the sourced scalar and tensor spectra and discuss the resulting modification of the tensor-to-scalar ratio. Finally, the main conclusions and possible extensions are summarized in Sec. \ref{conclusion}.

\section{\label{spectra}Production of fermion pairs during inflation}

\subsection{\label{model}The Dynamic Equations} 

The exponential expansion of the early universe is driven by a single scalar field that can subsequently decay into various fundamental particles.
In this work we investigate the production of fermions during inflation through a Yukawa interaction in which the fermions couple directly to the inflaton. The total Lagrangian is given by
\begin{align}
    \mathcal{L}=\mathcal{L}_\phi+\mathcal{L}_D+\mathcal{L}_\text{Yukawa}. 
\end{align}
with the individual contributions \cite{Dirac1,Dirac2}
\begin{subequations} \label{L}
\begin{align}
    &\mathcal{L}_\phi/\sqrt{-g_{\mu\nu}}=\frac12\partial_\mu\phi\partial^\mu\phi-\frac12 m_\phi\phi^2, \label{L_phi}\\
    &\mathcal{L}_\psi/\sqrt{-g_{\mu\nu}}=\frac{\mathrm i}{2}[\bar{\psi} \bar{\gamma}^\mu D_\mu\psi-(D_\mu\bar{\psi})\bar{\gamma}^\mu\psi]-m_\psi\bar{\psi}\psi, \label{L_D}\\
    &\mathcal{L}_\text{Yukawa}/\sqrt{-g_{\mu\nu}}=-g\phi\bar{\psi}\psi. \label{L_Y}
\end{align}
\end{subequations}
Here $\mathcal{L}_\phi$ is the Lagrangian of the scalar inflaton $\phi$ that drives cosmic inflation, and $m_\phi$ is its mass. 
The second term, $\mathcal{L}_\psi$, describes the Dirac fermions: $\psi$ is the Dirac spinor, $\bar\psi=\psi^\dagger\gamma^0$ denotes its adjoint, and $m_\psi$ the represents static fermion mass. 
The two sectors interact via the Yukawa term $\mathcal{L}_\text{Yukawa}=-g\phi\bar{\psi}\psi$, where $g$ is the dimensionless coupling strength.
The line element for the Friedmann-Robertson-Walker universe with tensor perturbation is given by
\begin{align}
    \mathrm{d}s^{2}&=\mathrm{d}t^2-a^2 h_{ij} \mathrm{d}x^i\mathrm{d}x^j \nonumber \\
    &=a^2{\left(\mathrm{d}\tau^2-\delta_{ij}\mathrm{d}x^i\mathrm{d}x^j\right)},  \label{metric}
\end{align}
with $\mathrm{d}t=a\mathrm{d}\tau $, where $t$ is the cosmic time, $\tau$ is the conformal time, and $a$ is the cosmic scale factor.
For exponentially expanding universe, the scalar factor can be expressed as $a=-1/H\tau$. 
The covariant derivatives appearing in $\mathcal{L}_D$ are $D_\mu\psi=\partial_\mu\psi+\Gamma_\mu\psi$ and $D_\mu\bar\psi=\partial_\mu\bar\psi-\bar\psi\Gamma_\mu$, 
where the spin connection in curved spacetime satisfies $\Gamma_0=0$ and $\Gamma_i=\frac{\dot a}{2}\gamma^0\gamma^i$, with $\dot a=\mathrm d a/\mathrm d t$ \cite{Dirac1,Dirac2}. 
In a Friedmann–Robertson–Walker universe, the curved-space gamma matrices are related to the flat-space ones by $\bar\gamma^0=-\gamma^0$, $\bar\gamma^i=a^{-1}\gamma^i$. 
Their spatial contraction yields \cite{Dirac3} $\bar\gamma^i\Gamma_i=\frac32 H\gamma^0$, where $H$ represents the Hubble parameter.

In warm inflationary scenario, the dissipative term $\Upsilon\dot\phi$ is the linear response of $g\bar\psi\psi$,
thus the equation of motion of scalar field perturbed from the thermal fluctuation noise $\xi$ in configure coordinate reads 
\begin{align}
    \partial_{\mu} \partial^{\mu} \phi-m_\phi \phi= \Upsilon u^{\mu} \partial_{\mu} \phi + \xi.  \label{Eq_phi}
\end{align}
Here the expressions of dissipative term at high temperature regime is given by \cite{dissipative1}
\begin{align}
    \Upsilon = 7.3 \frac{T^2}{g^2 m_{\phi}}\left[\frac{m_{\phi}}{m_{\phi}(T)}\right]^3, \label{Upsilon}
\end{align}
where $T$ denotes the cosmic temperature, and $m_{\phi}^2(T) \approx m_{\phi}^2 + \frac{g^2 T^2}{12}$ is the total effective mass.
In addition, the fluctuation term in Eq. \eqref{Eq_phi} is defined as  \cite{shape}
\begin{align}
    \xi=g(\bar{\psi} \psi-\langle \bar{\psi}\psi \rangle). 
\end{align}
Therefore it follows the dissipative-fluctuation relation in the long-wavelength description
\begin{align}
    \langle \xi(x) \xi(x') \rangle = 2 \Upsilon T a^{-3} \delta^4(x - x'). \label{dissipative-fluctuation-x}
\end{align}
with symbol $\langle\cdots\rangle$ representing the ensemble average. 
We note that the dissipative coefficient \eqref{Upsilon} is the approximate expression with high temperature limit, i.e., $m_\phi \ll T$, 
since, as discussed later, only the high temperature limit or relativistic limit predicts additionally interesting phenomena in the frame of warm inflation; 
otherwise, it degenerates to the general warm inflation with single scalar field.
Thus, the condition at low temperature regime, as seen in Refs. \cite{dissipative1,dissipative2}, does not take into consideration in this work.

The thermalized Dirac field is treated as a component of the thermal bath, which is typically modelled as an ideal fluid. Conservation of total energy and momentum then yields
\begin{align}
    \partial_{\mu}  T_{D\ \nu}^{\mu} = -\partial_{\nu} \phi ( \Upsilon u^{\mu} \partial_{\mu} \phi +\xi ). \label{EM_D}
\end{align}
The energy-momentum tensor of the Dirac fluid takes the form
\begin{subequations} \label{T_D}
\begin{align}
    &T_{D\,0}^{0} = -\rho_{D0}(1 + \mathcal E), \\
    &\partial_i T^i_{D\,0} = -(1+w) \frac{\rho_{D0}}{a^2} \nabla^2 \Theta,\\
    &T_{Dj}^{i} = (p_{D0} + \delta p_D) \delta_{j}^{i} = \rho_{D0}(w + c_{s}^{2} \mathcal E) \delta_{j}^{i}.
\end{align}
\end{subequations}
Here $\rho_{D0}$ and $p_{D0}$ denote the background energy density and pressure of the Dirac fluid, $w\equiv p_{D0}/\rho_{D0}$ is the equation-of-state parameter, and $c_s^2\equiv\delta p_D/\delta\rho_D$ is the sound speed.
At leading order, Eq.~\eqref{EM_D} reduces to
\begin{align}
    \dot{\rho}_{D0}+3H(\rho_{D0}+p_{D0})=\Upsilon\dot{\phi}_{0}^2,
\end{align}
where a dot denotes differentiation with respect to cosmic time $t$. For a (quasi-)equilibrium configuration this simplifies further to
\begin{align}
    \rho_{D0}\approx\frac{Q}{1+w}\dot{\phi}_{0}^2, \label{leading_EM_D}
\end{align}
with $Q=\Upsilon/(3H)$.
We expand the scalar field to first order as
\begin{align}
    \phi=\phi_{0}+\dot{\phi}_{0}\Phi, \label{phi}
\end{align}
where $\phi_0=\langle\phi\rangle$ is the homogeneous background. 
By substituting Eqs.~\eqref{T_D} and \eqref{phi} into the continuity equations \eqref{EM_D} (for both $\nu=0$ and $\nu=i$) as well as into the equation of motion of inflaton \eqref{Eq_phi}, 
making use of the background relation \eqref{leading_EM_D}, and neglecting the slow-roll term proportional to $(m_\phi/H)^2$, 
it yoelds the following linearized system in momentum space, expressed in terms of conformal time $\tau=-1/(aH)$:
\begin{subequations} \label{equations_1}
\begin{align}
    &\frac{\mathrm{d}\Theta}{\mathrm{d}\tau}-\frac{2}{\tau}\Theta-\frac{c^{2}_{s}}{1+w}\frac{\mathcal E}{H\tau}-\frac{3}{\tau}\Phi=0, \\
    &\frac{\mathrm{d}\mathcal E}{\mathrm{d}\tau}-\frac{\mathcal E}{\tau}+(1+w)k^{2}H\tau\Theta=-\frac{3}{\Upsilon\dot{\phi}_{0}\tau}\xi,\\
    &\frac{\mathrm{d}^2\Phi}{\mathrm{d}\tau^2}-\frac{2+3Q}{\tau}\frac{\mathrm{d}\Phi}{\mathrm{d}\tau}+\frac{\Upsilon}{(1+w)H^{2}\tau^{2}}\mathcal E+k^{2}\Phi=\frac{\xi}{\dot{\phi}_{0}H^{2}\tau^{2}}.
\end{align}
\end{subequations}
Consequently, the dissipative-fluctuation relation \eqref{dissipative-fluctuation-x} becomes, in momentum space,
\begin{align}
    \langle \xi(\mathbf{k},\tau) \xi^*(\mathbf{k}',\tau') \rangle = 
        2 (2\pi)^3 \Upsilon T H^4 z^4 \frac{1}{k^{3}} \delta^3(\mathbf{k} - \mathbf{k}') \delta(z - z'). \label{dissipative-fluctuation-k}
\end{align}

\section{Power spectrum of scalar perturbation \label{P_large}}

Define the new variable $z\equiv-k\tau$, ranging from 0 to $\infty$. Then the dynamical equations \eqref{equations_1} become the form as a function of $z$:
\begin{subequations} \label{equations_2}
\begin{align}
	&\Theta' - \frac{2}{z} \Theta + \frac{c_s^2}{1 + w } \frac{\mathcal{E}}{Hz} - \frac{3}{z} \Phi = 0, \label{Theta_z}\\
	&\mathcal{E}' - \frac{1}{z} \mathcal{E} + (1 + w)Hz \Theta = -\frac{\xi}{\Upsilon \dot{\phi}_0 z}, \label{E_z}\\
    &\Phi'' - \frac{3 + 2Q}{z} \Phi' + \dfrac{\Upsilon}{(1 + w)H z^2} \frac{\mathcal{E}}{H} + k^2 \Phi = \frac{1}{\dot{\phi}_0 H^2 z^2}\xi, \label{Phi_z}
\end{align} 
\end{subequations}
where the prime $'$ denotes the derivative with respect to $z$. 
By eliminating the variables $\Theta$ and $\mathcal{E}$ in Eq. \eqref{Phi_z}, we obtain the equation of motion about $\Phi$ in terms of a Volterra integral equation of the second kind
\begin{align}
	&\Phi(z) = \frac{1}{\dot{\phi}_{0} H^2} \int \frac{dz_1}{z_1^2} G_k^{(\Phi)}(z, z_1) \xi(z_1) \nonumber\\
	&+ \frac{9Qc_s^2}{1 + \omega} \frac{1}{\Upsilon \dot{\phi}_{0} H} \int \frac{dz_1}{z_1^2} G_k^{(\Phi)}(z, z_1) \int dz_2 G_{c_s, k}^{(\Psi)}(z_1, z_2) \xi(z_2) \nonumber\\
	&+ 9Qc_s^2 \int \frac{dz_1}{z_1^2} G_k^{(\Phi)}(z, z_1) \int dz_2 G_{c_s, k}^{(\Psi)}(z_1, z_2) \Phi(z_2).  \label{Phi_int}
\end{align}
The derivations of the equation are presented in detail in Appendix \ref{Dev_Phi}.
Moreover, the Green’s functions $G_{c_s,k}^{(\Psi)}$ and $G_k^{(\Phi)}$ are given in closed form in Eqs. \eqref{G_Psi} and \eqref{G_Phi}, respectively.
Each term on the right-hand side of Eq. \eqref{Phi_int} carries a specific physical interpretation:
the first term represents the direct perturbation arising from the thermal fluctuation $\xi$, corresponding to the scalar perturbation generated by thermal noise;
the second term accounts for the perturbation induced by thermal–acoustic mixed scattering (i.e., the propagation of thermal fluctuations among the fermions), thereby describing the coupling between heat and sound;
the third term comprises the iterative contributions of the two preceding processes and therefore represents higher-order scattering effects arising from the interplay of thermal and acoustic fluctuations.

\begin{figure*}
	\center
	\includegraphics[width=\linewidth]{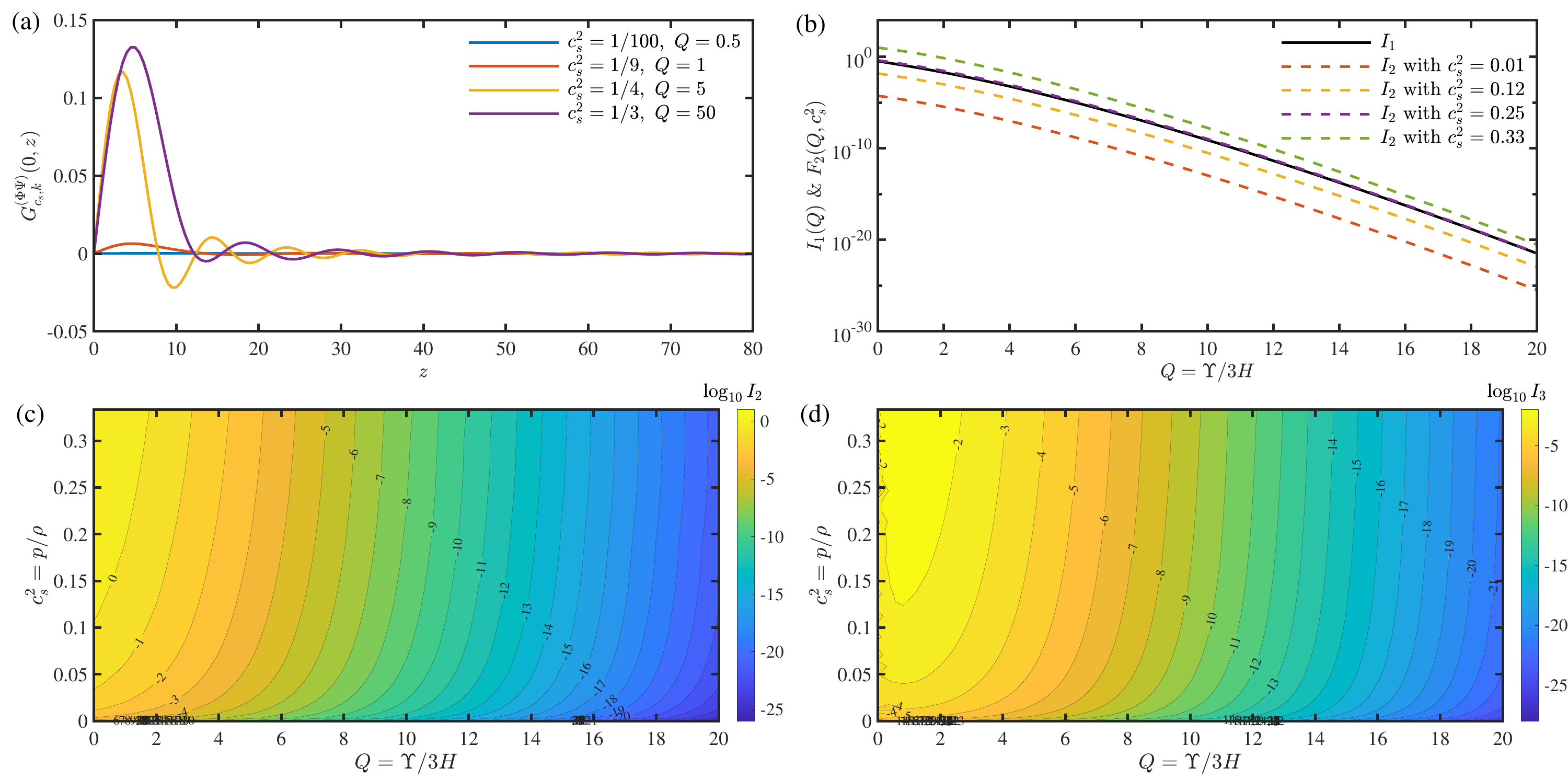}
	\caption{(a) Green's function $G_{c_s,k}^{(\Phi\Psi)}(0, z)$ as a function of $z$; (b) Amplitude of $I_1(Q)$ and $I_2(Q,c_s^2)$ with different values of $c_s^2$ as a function of $Q$; 
            (c) Contour of logarithm of $I_2(Q,c_s^2)$ in the $Q$-$c_s^2$ plane; (d) Contour of logarithm of $I_3(Q,c_s^2)$ in the $Q$-$c_s^2$ plane. }
	\label{G_F_123}
\end{figure*}

Applying the successive approximation method, the solution to Eq. \eqref{Phi_int} decomposes into the terms as follow \cite{successive1,successive2}:
\begin{align}
	\Phi = \Phi_0 + 9Qc_s^2\Phi_1 + \left(9Qc_s^2\right)^2\Phi_2 + \dots,
\end{align}
where 
\begin{subequations} \label{equations_3}
\begin{align}
	\Phi_0 = &\frac{3c_s^2}{1+w} \frac{1}{\dot{\phi}_{0} H^2} \int \mathrm{d}z_2 \, G_{c_s, k}^{(\Phi \Psi)}(z, z_2) \xi(z_2) \nonumber\\
        &+ \frac{1}{\dot{\phi}_{0} H^2} \int \mathrm{d}z_1 \, G_k^{(\Phi)}(z, z_1) \xi(z_1), \\
	\Phi_1 = &\int \mathrm{d}z_2 \ G_{c_s,k}^{(\Phi \Psi)}(z, z_2) \Phi_0(z_2), \\
	\Phi_2 = &\int \mathrm{d}z_2 \, G_{c_s, k}^{(\Phi\Psi)}(z, z_2) \Phi_1(z_2), \\
		&\qquad \cdots \notag
\end{align}
\end{subequations}
For convenience, we introduce a new Green's function in the above equations 
\begin{align}
	G_{c_s,k}^{(\Phi\Psi)}(z, z_2) = \int \mathrm{d}z_1 \, z_1^{-2} G_k^{(\Phi)}(z, z_1) G_{c_s,k}^{(\Psi)}(z_1, z_2).  \label{G_Phi_Psi}
\end{align}
Figure \ref{G_F_123}(a) displays the curves of $G_{c_s,k}^{(\Phi\Psi)}(0,z)$ versus $z$. These plots make it evident that $|G_{c_s,k}^{(\Phi\Psi)}|\ll 1$ holds for all values of $Q$ and $c_s^2$. 
Consequently, the first-order solution for $\Phi$ takes the form
\begin{align}
	&\Phi \approx \frac{1}{\dot{\phi}_0 H^2} \int \mathrm{d}z_1 \, z_1^{-2} G_k^{(\Phi)}(z, z_1) \xi(z_1) \nonumber\\
    &+ \frac{9Q c_s^2}{1 + w} \frac{1}{\Upsilon \dot{\phi}_0 H} \int \mathrm{d}z_1 \, G_{c_s, k}^{(\Phi \Psi)}(z, z_1) \xi(z_1) \nonumber\\
    &+ \frac{9Q c_s^2}{\dot{\phi}_0 H^2} \int \mathrm{d}z_1 \, G_{c_s, k}^{(\Phi \Psi)}(z, z_1) \int \mathrm{d}z_2 \, z_2^{-2} G_{k}^{(\Phi)}(z_1, z_2) \xi(z_2). \label{Phi_sol}
\end{align}
The first term on the right-hand side describes pure thermal propagation, the second term corresponds to acoustic–thermal propagation, and the third term accounts for thermal–acoustic–thermal propagation.

The Green's function $G_{c_s,k}^{(\Phi\Psi)}$ denotes the propagation of thermal fluctuation scattered by the Dirac field. 
As shown in Fig. \ref{G_F_123}(a), the amplitude of this propagation essentially vanishes in the non-relativistic regime ($c_s^2\sim0$).
In contrast, when $c_s^2\lesssim1/3$ (the relativistic regime), a clear propagation signature appears.
This behaviour can be interpreted as a competition between the self-gravity of the fermion particles and Pauli blocking:
the fermions produced by inflaton decay generate a gravitational attraction, while the Pauli exclusion principle, which is an intrinsic property of fermions, gives rise to a positive degeneracy pressure.
Only when these two effects become comparable (i.e., in the relativistic limit) propagates an acoustic wave through the Dirac fields.
If the acoustic propagation is absent, fermion production is sufficiently weak that the scalar perturbation arises solely from the thermal fluctuation $\xi$, corresponding to the first term on the right-hand side of Eq. \eqref{Phi_sol}.

In the spatially flat gauge, the comoving curvature perturbation relevant for observations is defined as $\mathcal{R}=H\delta\phi/\dot\phi_0=H\Phi$.
Therefore the two-point correlation function of the scalar perturbation in the super-horizon limit is given by
\begin{align}
	& \langle \mathcal{R}(\mathbf{k}_1, \tau) R(\mathbf{k}_2, \tau) \rangle \bigg|_{k_1, k_2 \ll a_* H_*} \nonumber \\
	&\approx \lim_{-k_1 \tau \to 0} \lim_{-k_2 \tau \to 0} \left[ \frac{1}{(\dot{\phi}_0 H)^2} \int_{\infty}^{z_1} \mathrm{d}z_1' \int_{\infty}^{z_2} \mathrm{d}z_2' G_{k_1}^{(\Phi)}(z_1, z_1') G_{k_2}^{(\Phi)}(z_2, z_2') z_1'^{-2} z_2'^{-2} 
        \left\langle \xi(\mathbf{k}_1, z_1') \xi(\mathbf{k}_2, z_2') \right\rangle \right. \nonumber \\
	&\quad + \left( \frac{9 Q c_s^2}{1 + w} \frac{1}{\Upsilon \dot{\phi}_0 H} \right)^2 \int_{\infty}^{z_1} \mathrm{d}z_1' \int_{\infty}^{z_2} \mathrm{d}z_2' G_{c_s,k_1}^{(\Phi \Psi)}(z_1, z_1') G_{c_s, k_2}^{(\Phi \Psi)}(z_2, z_2')
         \left\langle \xi(\mathbf{k}_1, z_1') \xi(\mathbf{k}_2, z_2') \right\rangle  \nonumber \\
	&\quad \left. + \frac{9 Q c_s^2}{1 + w} \frac{1}{\Upsilon \dot{\phi}_0^2 H^3} \int_{\infty}^{z_1} \mathrm{d}z_1' \int_{\infty}^{z_2} \mathrm{d}z_2' G_{c_s, k_1}^{(\Phi \Psi)}(z_1, z_1') G_{k_2}^{(\Phi )}(z_2, z_2') z_2'^{-2} 
        \left\langle \xi(\mathbf{k}_1, z_1') \xi(\mathbf{k}_2, z_2') \right\rangle + (\mathbf k_1 \leftrightarrow \mathbf k_2) \right], \label{cor_R_1}
\end{align}
where $z_{1,2}=-k\tau_{1,2}$ and $(\mathbf{k}_1\leftrightarrow\mathbf{k}_2)$ denotes the interchange of $\mathbf{k}_1$ and $\mathbf{k}_2$ relative to the preceding expression.
The first and second terms on the right-hand side of Eq.~\eqref{cor_R_1} correspond, respectively, to the auto-correlations of the pure thermal contribution (the integral involving $G^{(\Phi)}$) 
and of the thermal–acoustic mixed contribution (the integral involving $G^{(\Phi\Psi)}$), both of which appear on the right-hand side of Eq.~\eqref{Phi_sol}; the third term is their cross-correlation.
Higher-order contributions involving more than two Green’s functions have been omitted. This truncation is justified by the rapid decrease in amplitude as the number of Green’s functions increases, as will be discussed shortly. 
Note that the super-horizon limit corresponds to $z\to0$ inside the Green’s function $G_k^{(\Phi)}(z,z')$, with exact expression \eqref{G_Phi}.
Moreover, the small-argument asymptotics of the Bessel functions, $\mathrm{J}_\nu(z)\approx0$ and $\mathrm{N}_\nu(z)\approx-\frac{\Gamma(\nu)}{\pi}\bigl(\frac{2}{z}\bigr)^\nu$ for $z\to0$ and $\nu>0$, reduce the Green’s function to
\begin{align}
	G_{k}^{(\Phi)}(z, z')|_{z\to0}\approx\frac{2^{\nu-1}\Gamma(\nu)}{z'^{\nu-1}}\mathrm J_\nu(z')\theta(z'), 
\end{align}
where $\Gamma(z)$ denotes the Gamma function and the remaining mathematical symbols are reffered in Eq.~\eqref{G_Phi}. The same approximation is employed for the numerical results shown in Fig.~\ref{G_F_123}(a).
Applying the fluctuation–dissipation relation \eqref{dissipative-fluctuation-k}, the two-point correlation function then becomes
\begin{align}
	\langle \mathcal{R}(\mathbf{k}_1, 0) R(\mathbf{k}_2, 0) \rangle = &\frac{16 \pi^5}{k^3} \delta^3(\mathbf{k}_1 + \mathbf{k}_2) \left( \frac{H^2}{2\pi\dot{\phi}_0^2} \right)^2
        \frac{T}{H} \frac{\Upsilon}{H} \left[ 2^{\nu-1} \Gamma(\nu) \right]^2 \cdot  \nonumber \\
    &\left[I_1(\nu) + \left( \frac{3 c_s^2}{1 + w} \right)^2 I_2(\nu,c_s^2) + \frac{6 c_s^2}{1 + w} I_3(\nu,c_s^2)  \right],  \label{cor_R_2}
\end{align}
with $Q=\Upsilon/3H$ and $\nu=\frac{Q}{2}+\frac32$. Here, the first parameter $I_1$ is expressed as
\begin{align}
	I_1(\nu)&=\int_0^\infty \mathrm{d}z \, z^{2-2\nu}\mathrm J_\nu^2(z) 
    = \frac{\Gamma(\nu-1)\Gamma(3/2)}{2\sqrt{\pi}\Gamma(\nu-1/2)\Gamma(2\nu-1/2)}.
\end{align}
The second parameter is 
\begin{align}
    I_2(\nu,c_s^2)  &=\int_0^\infty \mathrm{d}z_1 \int_0^\infty \mathrm{d}z_2 \frac{\mathrm{J}_\nu(z_1) \mathrm{J}_\nu(z_2)}{z_1^{\nu} z_2^{\nu}} \int_{\max(z_1, z_2)}^{\infty} \frac{\mathrm{d}z}{z^2}\cdot  \nonumber \\
	&\quad\quad \left[ (1 + c_s^2 z_1 z) \sin c_s(z_1 - z) + c_s(z - z_1) \cos c_s(z_1 - z) \right]\cdot  \nonumber \\
	&\quad\quad \left[ (1 + c_s^2 z_2 z) \sin c_s(z_2 - z) + c_s(z - z_2) \cos c_s(z_2 - z) \right]  \nonumber \\
    &= 2 \int_0^\infty \mathrm{d}z \int_0^z \mathrm{d}z_1 \int_0^{z_1} \mathrm{d}z_2 \, \frac{\mathrm{J}_\nu(z_1) }{z_1^\nu} \frac{\mathrm{J}_\nu(z_2)}{z_2^\nu} \frac{1}{z^2}\cdot \nonumber \\
	&\quad\quad \left[ (1 + c_s^2 z_1 z) \sin c_s (z_1 - z) + c_s (z - z_1) \cos c_s (z_1 - z) \right]\cdot  \nonumber \\
	&\quad\quad \left[ (1 + c_s^2 z_2 z) \sin c_s (z_2 - z) + c_s (z - z_2) \cos c_s (z_2 - z) \right],
\end{align}
where the factor 2 appearing in the second equality arises from the permutation of $z_1$ and $z_2$ in the integral lower-limit $\max(z_1, z_2)$. 
Similarly, the third reads 
\begin{align}
    &I_3(\nu,c_s^2) \nonumber \\ &= 2\int_0^\infty \mathrm{d}z_1 \int_{z_1}^\infty \mathrm{d}z \, \frac{\mathrm{J}_\nu(z_1) \mathrm{J}_\nu(z)}{z_1^{\nu} z^{\nu}}  
	   \left[( 1 + c_s^2 z z_1 ) \sin c_s (z_1 - z) + c_s (z - z_1) \cos c_s (z_1 - z) \right].
\end{align}

In Fig.~\ref{G_F_123}(b) we present the numerical results for $I_1$ (solid curve) as a function of $Q$, together with the corresponding curves of $I_2(Q,c_s^2)$ (dashed) for several values of $c_s^2$. 
Interestingly, the curves of $I_2$ remain nearly parallel to that of $I_1$, even though the different values of $c_s^2$ are clearly distinguishable.
Panels (c) and (d) of the same figure present contour plots of $\log_{10} I_2$ and $\log_{10} I_3$ in the $Q$–$c_s^2$ plane.
In the non-relativistic regime, the amplitudes of both integrals are strongly suppressed and essentially vanish, whereas acoustic propagation yields a significant contribution to the correlation function \eqref{cor_R_2}.
These findings are fully consistent with the earlier analysis of the Green’s function $G^{(\Phi\Psi)}$.
Moreover, for any fixed $Q$ and $c_s^2$ the magnitude of $I_2$ is roughly two orders of magnitude larger than that of $I_3$.
Consequently, integrals that involve additional Green’s functions decrease rapidly, confirming the neglect of the higher-order terms that appear in the auto-correlation function \eqref{cor_R_1}.
Because the non-relativistic limit simply recovers the standard single-field warm-inflation result, the subsequent analysis focuses on the relativistic regime ($c_s^2\lesssim1/3$), where interesting cosmological signatures may arise.

\section{Particle Density and Acoustic Velocity of Thermalized Fermion \label{sec_nf}}

The ratio of the cosmic temperature $T$ to the inflationary Hubble parameter $H$ plays a central role in our analysis, as is already evident from the preceding calculations.
This ratio is necessarily related to the fundamental parameters of the model.
Because the energy of the fermions generates from the inflaton potential and the Dirac particles behave as a relativistic fluid, energy conservation yields
\begin{align}
    \Upsilon\dot\phi_0^2=\frac{\pi^2}{30}N_c T^4.
\end{align}
Here $N_c=3$ is the number of quark degrees of freedom.
The Friedmann equation of warm inflation further implies $3H(1+Q)\dot\phi_0+m_\phi^2 \phi\approx0$.
In the relativistic regime one has $Q\gg1$ and $m_\phi^2(T)\approx g^2T^2/12$, which together lead to the approximate relation
\begin{align}
    \frac{T}{H}\approx0.0942\,g\left(\frac{m_\phi}{H}\right)^{2/3}\left(\frac{m_\text{eff}}{H}\right)^{2/3}.   \label{TH}
\end{align}
Equation~\eqref{TH} will be employed throughout the subsequent analysis to estimate the temperature-to-Hubble ratio.

\subsection{Particle Density}

In quasi-de Sitter spacetime, the exact solutions of the Dirac fields for the chiral states $h=\pm$ and spin states $s=\pm$ are given in Eq.~\eqref{sol_psi_J}, which are expressed in terms of Bessel functions $\mathrm{J}_\nu(z)$.
A detailed derivation of these solutions can be found in Refs.~\cite{Dirac3,Dirac4}.
Using the corresponding eigenvectors, the quantized Dirac field $\hat\psi$ is written as in Eq.~\eqref{qunta_D}, where $\hat b$ denotes the particle annihilation operator and $\hat d^\dagger$ the antiparticle creation operator.
The net fermion number density in configuration space is then defined by $n_\psi(x)\equiv\langle\hat\psi^\dagger(x)\hat\psi(x)\rangle$.
Its explicit expression appears in Eq.~\eqref{n_psi_1} as a momentum-space integral.
Within the slow-roll approximation, a first-order Taylor expansion of $n_\psi$ in the parameter $\tilde\eta$ yields Eq.~\eqref{n_psi_2}.

The first interesting feature is the background net density $n_0=n_\psi|_{\tilde\eta=0}$, which corresponds to the fermion density in exact de Sitter spacetime ($\phi=\text{const}$ and $\dot H=0$). 
In this case, one has $\tilde\xi=\tilde m$ with $\tilde m=(m_\psi+g\phi)/H$. The analytic expression for $n_0$ is given in Eq.~\eqref{n0_app} as a momentum-space integral. Numerical simulation yields the compact result
\begin{align}
	n_0 = 0.3206\,(2\pi H)^3\,\tilde{m}\left(\frac{T}{H}\right)^2, \label{n0_num}
\end{align}
which has been verified to high accuracy. In deriving this expression, the chemical potential $\mu$ is identified with the effective mass $m_\text{eff}=m_\psi+g\phi$, as appropriate for an ideal relativistic fluid.
Equation~\eqref{n0_num} indicates that three ingredients must act simultaneously to produce fermions from the vacuum: spacetime geometric dynamics, a non-vanishing mass (chiral asymmetry), and a thermal bath. 
Gravitational particle production is generated by the inflationary expansion, quantified by the Hubble parameter $H$. 
Then, as shown in Ref.~\cite{production_m}, massless fermions experience no gravitational production in an exact Friedmann–Robertson–Walker geometry, therefore a non-zero mass is therefore essential. 
The resulting chiral asymmetry then yields a net density of observable fermions, in full agreement with~\eqref{n0_num}. 
Finally, the non-equilibrium thermal statistics associated with a finite cosmic temperature $T$ further contribute to the net fermion abundance.

\begin{figure}[h]
	\center
	\includegraphics[width=.5\linewidth]{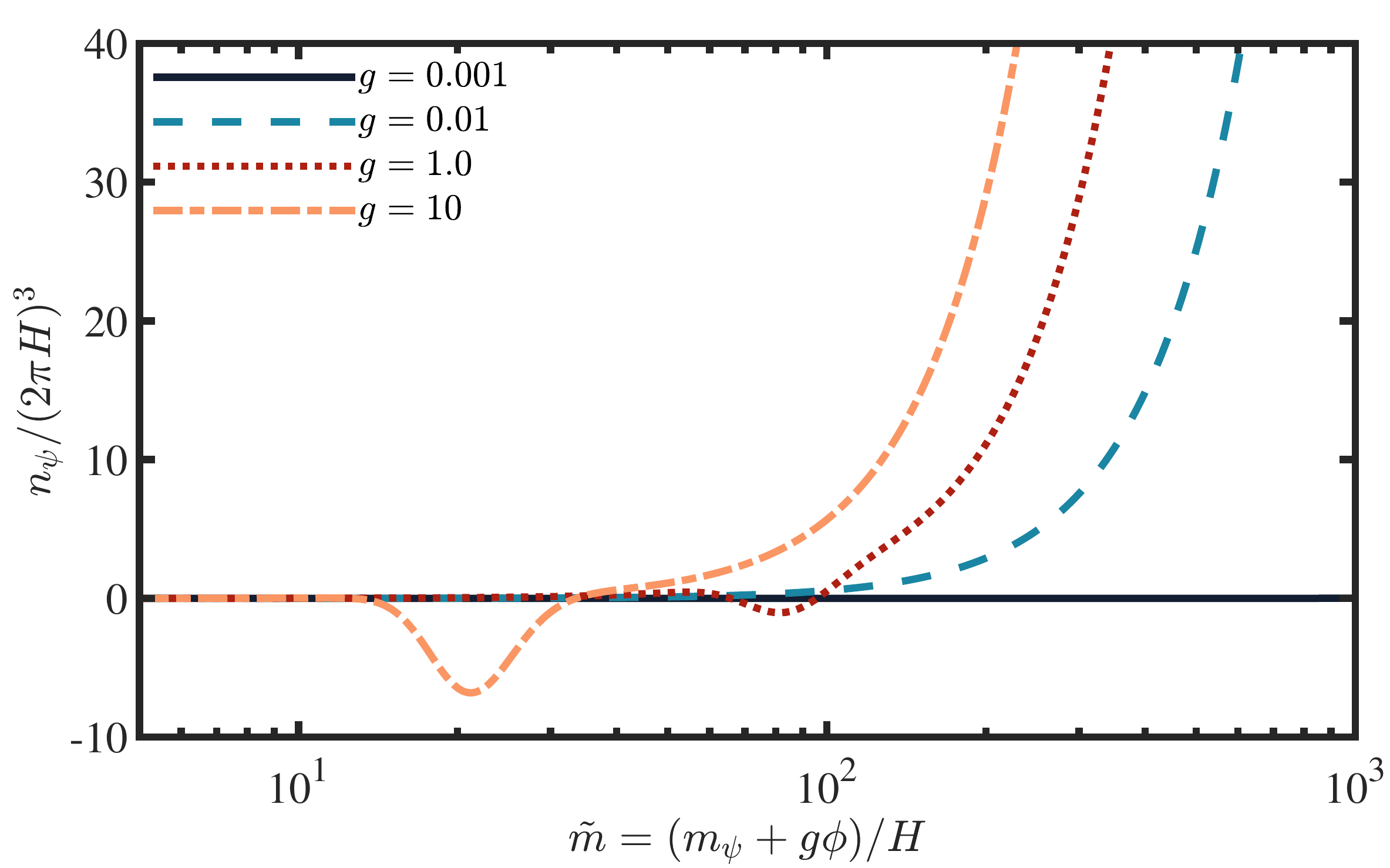}
	\caption{Curves of the first-order comoving net fermion density $n_\psi$ plotted against the dimensionless effective mass $\tilde{m}$ for several values of the Yukawa coupling $g$, which is in unit of $(2\pi H)^3$.
        For the case of $g=10$, the amplitude of $n_\psi$ is suppressed by roughly three orders of magnitude. }
	\label{n_psi_Fig}
\end{figure}

For the first-order correction to the net fermion density, Fig.~\ref{n_psi_Fig} shows the curves of $n_\psi$ as a function of the dimensionless effective mass $\tilde{m}$ for several values of the Yukawa coupling $g$.
When $g=10$, the amplitude is suppressed by roughly three orders of magnitude. In the numerical evaluation the scalar power spectrum is fixed at $\mathcal{P}_\zeta=(H^2/2\pi\dot\phi_0)\sim10^{-9}$.
The density grows rapidly with increasing $g$. For every value of the coupling, $n_\psi$ vanishes as $\tilde{m}\to0$, indicating that essentially massless fermions produce no appreciable net density within the present framework.
In contrast, a sharp rise sets in once $\tilde{m}\gtrsim10^{2}$. Physically, this occurs because a large effective mass modifies the fermionic mode functions in de Sitter spacetime and thereby enhances the non-adiabatic particle-production process.
In the limit $\tilde\eta\to0$, the first-order results coincide with the background density $n_0$, seen as in Eq.~\eqref{para} for definition of $\tilde\eta$.
When $g$ becomes sufficiently large, however, a clear departure from $n_0$ appears. Numerical evaluation shows that this departure follows a Gaussian profile arising from the series term on the right-hand side of Eq.~\eqref{n_psi_3}.
The same behaviour is independently confirmed by a numerical evaluation of the exact integral expression \eqref{n_psi_1}. 

It follows that $n_0$ evaluated at $\tilde{\eta}=0$ describes the adiabatic regime, in which the fermionic comoving density stays constant throughout the purely exponential expansion of the universe. 
For sufficiently weak Yukawa couplings ($g\ll 1$), the evolution of $n_\psi$ remains a near equivalence to the adiabatic case. 
Significant non-adiabatic effects appear once the coupling strength $g$ reaches or exceeds order unity, indicating strong non-adiabatic fermion production. 
The reason is that the Yukawa interaction $ -g\phi\bar{\psi}\psi$ directly couples the inflaton field to fermions, allowing energy transfer from the inflaton background into the fermion sector.
In this sense, the parameter $g$ itself serves as a quantitative indicator of the degree of non-adiabaticity about the generations of fermion. 

\subsection{Acoustic Velocity}

The acoustic velocity characterizes the relativistic properties of the fermion fluid.
For a perfect fluid with a linear barotropic equation of state $w$, the squared acoustic velocity is given by
\begin{align}
    c_s^2=w={p_\psi}/{\rho_\psi}, \label{cs2_w_equal}
\end{align}
where the energy density \(\rho_\psi\) and pressure \(p_\psi\) of the fermionic component are analytically derived in Eqs. \eqref{rho_app} and \eqref{p_psi_app}, respectively. 
The detailed derivations and related discussions are presented in Appendix \ref{rho_p_app}. In the following analysis, we continue to adopt the simplification $\tilde{\eta}=0$ in a de Sitter spacetime. 

\begin{figure}
	\center
	\includegraphics[width=.5\linewidth]{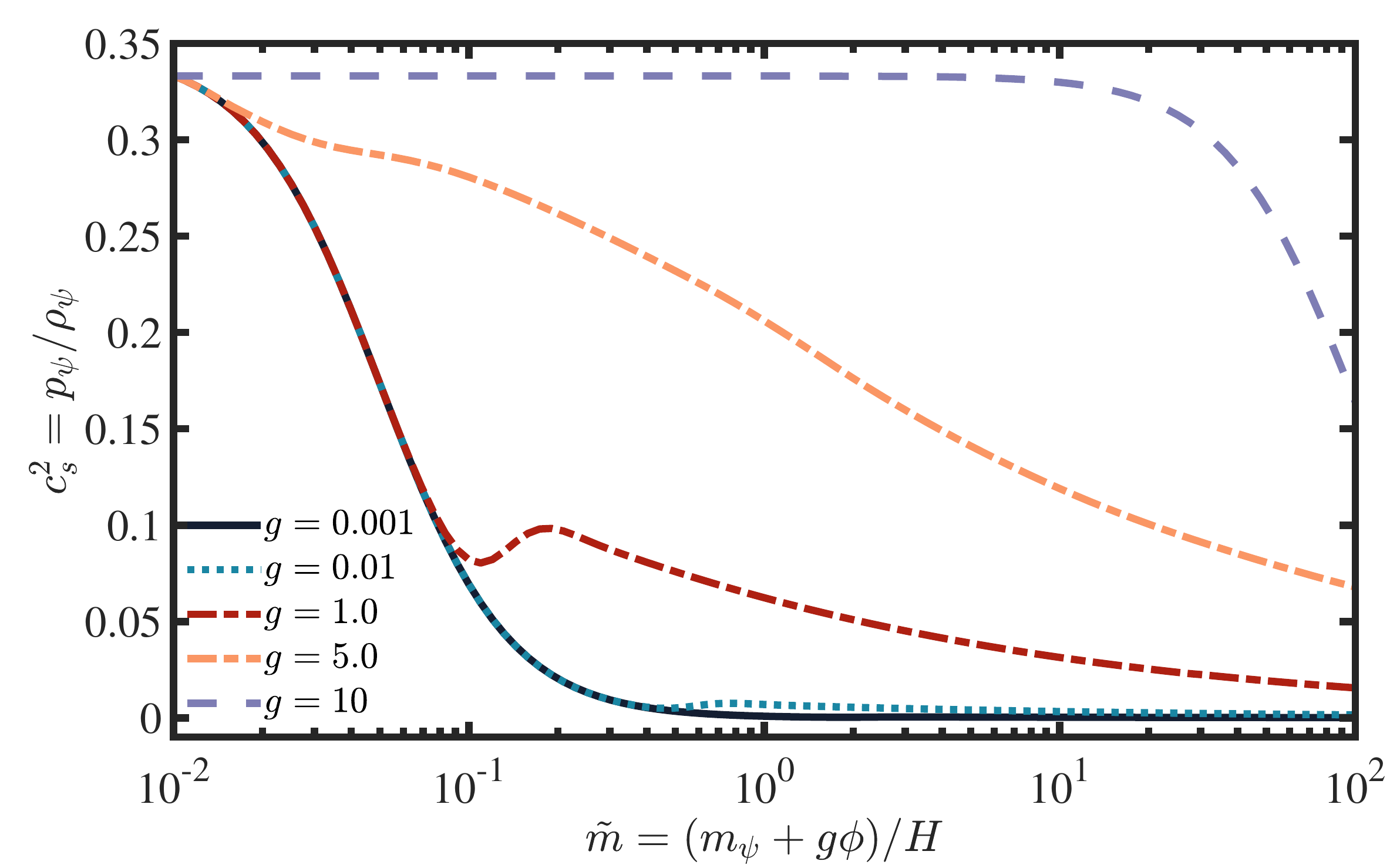}
	\caption{Curves of the fermionic acoustic velocity $c_s^2$ with respect to the dimensionless effective mass $\tilde m=(m_\psi+g\phi)/H$ for various values of Yukawa interaction strength $g$. }
	\label{cs2_Fig}
\end{figure}

In Figure \ref{cs2_Fig}, we plot the evolution of $c_s^2$ as a function of the dimensionless effective mass $\tilde{m}=(m_\psi+g\phi)/H$ for different values of the Yukawa coupling strength $g$. 
For sufficiently small coupling strength, such as $g=0.01$, the acoustic velocity decreases monotonically with increasing $\tilde{m}$. 
The curves corresponding to smaller values of $g$ almost overlap with this case, which is consistent with the behavior obtained in flat spacetime.

As $g$ increases, the corresponding curves begin to deviate from the weak-coupling case beyond a critical value of $\tilde{m}$. 
Interestingly, after the deviation occurs, an oscillatory feature emerges before the curves gradually return to a monotonic decreasing behavior. 
This phenomenon can be understood as the result of a competition between two effects. 
On the one hand, an increasing effective mass, $m_{\rm eff}=m_\psi+g\phi$, suppresses the relativistic nature of the fermion fluid. 
On the other hand, a larger effective mass also leads to an increase in the cosmic temperature $T$, which enhances the relativistic contribution, as described by Eq. \eqref{TH}. 
The interplay between these two competing mechanisms gives rise to the local oscillatory behavior.

For sufficiently large coupling strength ($g\gg1$), the fermion fluid eventually enters a non-relativistic regime when $\tilde{m}\gg1$. 
It is therefore evident that for weak Yukawa coupling, the fermion fluid can remain relativistic only within the regime of sufficiently small effective mass. 
In contrast, strong coupling allows the relativistic regime to persist over a much wider range of $\tilde{m}$. These results provide important implications for subsequent analyses of the tensor-to-scalar ratio.

\section{Primordial perturbations Reduced from thermalized Fermions \label{sec_GW}}

\subsection{Primordial Scalar and Tensor Spectra}

The spectral densities of scalar perturbations $\mathcal P_\mathcal{R}$ and tensor perturbations $\mathcal P_h$ are defined as
\begin{align}
    &\left\langle\hat{\mathcal{R}}(\mathbf{k})\hat{\mathcal{R}}^\dagger(\mathbf{k}^{\prime})\right\rangle=\frac{2\pi^2}{k^3}\delta^3(\mathbf{k}-\mathbf{k}^{\prime})\mathcal{P}_{\mathcal{R}}(k),\\
    &\sum_{\lambda=\pm}\left\langle\hat{h}_\lambda(\mathbf{k})\hat{h}_\lambda^\dagger(\mathbf{k}^{\prime})\right\rangle=\frac{2\pi^2}{k^3}\delta^3(\mathbf{k}-\mathbf{k}^{\prime})\mathcal{P}_h(k).
\end{align}
where $\langle\hat{\mathcal{R}}(\mathbf{k})\hat{\mathcal{R}}^\dagger(\mathbf{k}^{\prime})\rangle$ and $\langle\hat{h}_\lambda(\mathbf{k})\hat{h}_\lambda^\dagger(\mathbf{k}^{\prime})\rangle$ 
are the two-point correlation functions of scalar and tensor perturbation, respectively. 
Both perturbations have two contributions: one from vacuum and the other from the sourced Dirac field.
Since these two contributions are statistically independent, the total spectral densities can be written as $\mathcal{P}_{\mathcal{R}}=\mathcal{P}_{\mathcal{R}}^{(v)}+\mathcal{P}_{\mathcal{R}}^{(s)}$
and $\mathcal{P}_h=\mathcal{P}_h^{(v)}+\mathcal{P}_h^{(s)}$, with the standard vacuum spectra
\begin{align}
    \mathcal{P}_{\mathcal{R}}^{(v)}=\left(\frac{H^2}{2\pi\dot{\phi}_0}\right)^2,\quad \mathcal{P}_h^{(v)}=\frac{2H^2}{\pi^2M_p^2}.
\end{align}
The sourced spectra $\mathcal{P}_{\mathcal{R}}^{(s)}$ and $\mathcal{P}_h^{(s)}$ are obtained directly from Eqs. \eqref{cor_R_2} and \eqref{t_correl}. 
To quantize the spectrums reduced from the thermalized Dirac fields, we define the ratio of perturbed ones to the vacuum ones as the parameters to illustrate the magnitudes:
\begin{align}
    R_s=\frac{\mathcal P^{(s)}_s}{\mathcal P^{(v)}_s},\quad R_h=\frac{\mathcal P^{(s)}_h}{\mathcal P^{(v)}_h}. 
\end{align}

\begin{figure}
	\center
	\includegraphics[width=.5\linewidth]{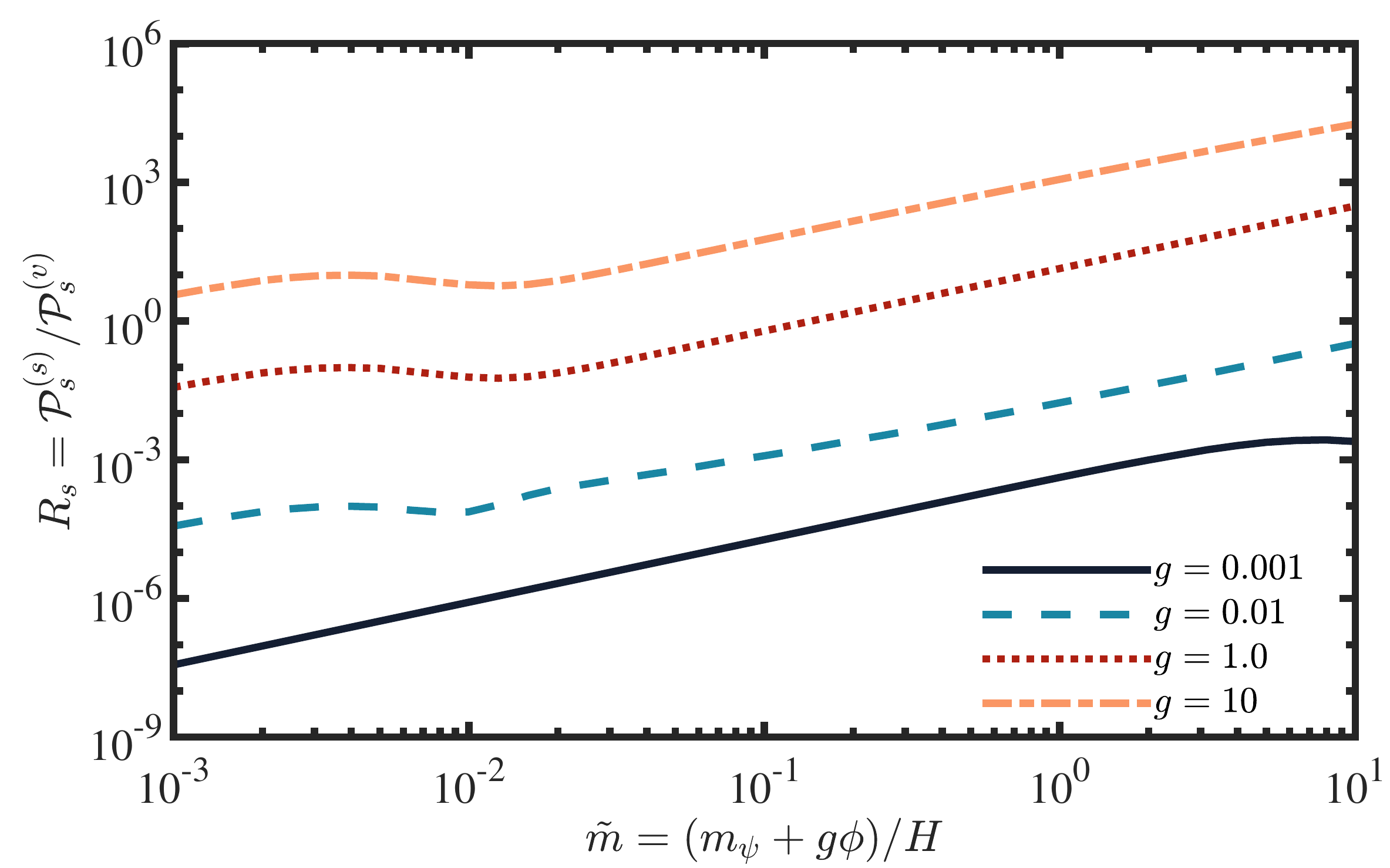}
	\caption{Magnitudes of sourced primordial scalar perturbation $R_s=\mathcal P^{(s)}_s/\mathcal P^{(v)}_s$ with respect to the dimensionless effective mass $\tilde m=(m_\psi+g\phi)/H$ for various values of Yukawa interaction strength $g$. }
	\label{R_s_Fig}
\end{figure}

\begin{figure}
	\center
	\includegraphics[width=.5\linewidth]{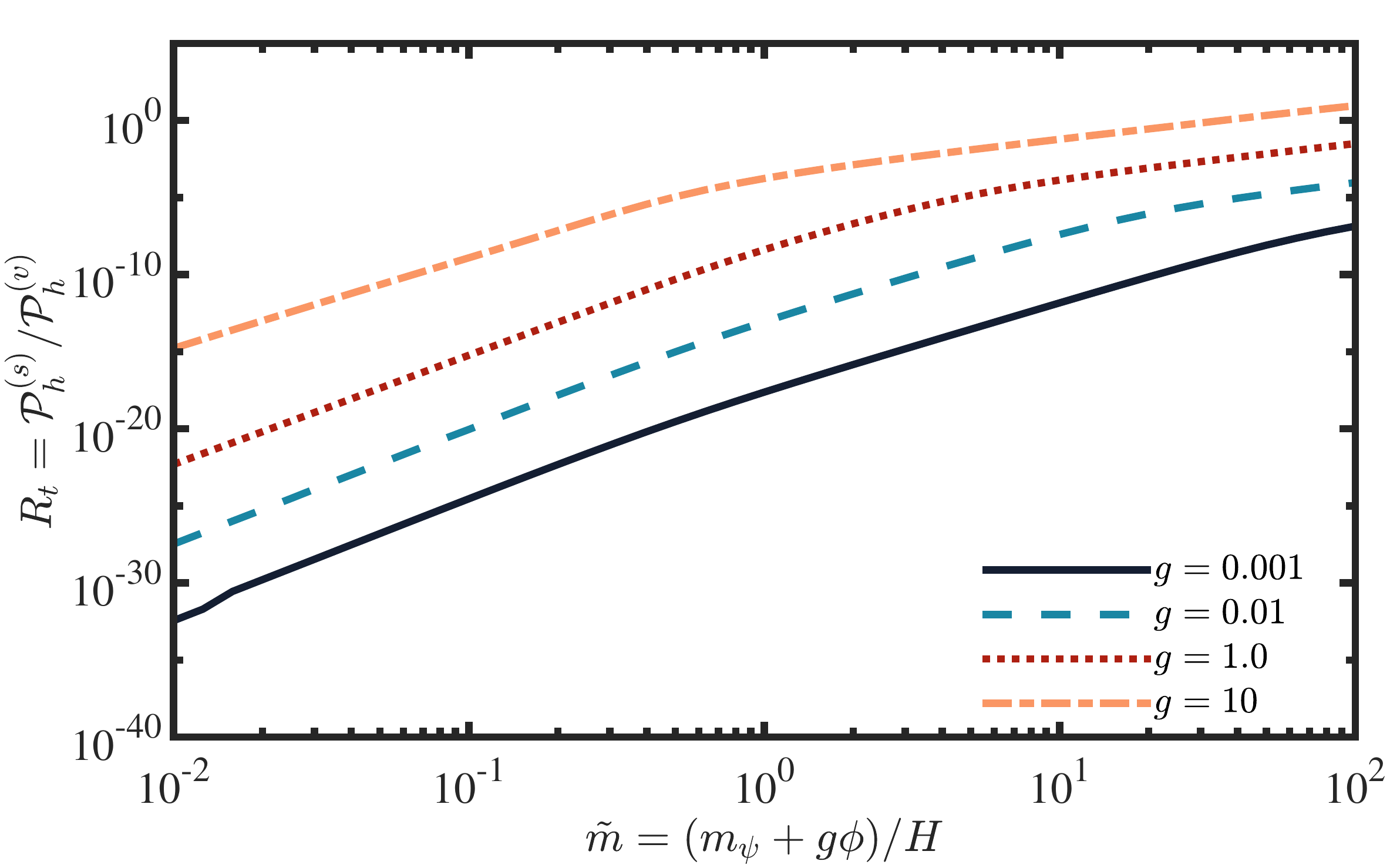}
	\caption{Magnitudes of sourced primordial tensor perturbation $R_h=\mathcal P^{(s)}_h/\mathcal P^{(v)}_h$ with respect to the dimensionless effective mass $\tilde m=(m_\psi+g\phi)/H$ for various values of Yukawa interaction strength $g$. }
	\label{R_h_Fig}
\end{figure}

The numerical results for $R_s$ and \(R_h\) are shown in Figs. \ref{R_s_Fig} and \ref{R_h_Fig}, respectively. 
As seen in Fig. 4, the sourced scalar perturbation is sensitive to both the Yukawa coupling strength $g$ and the dimensionless effective mass $\tilde m$. 
For weak coupling, $R_s\ll1$ over the considered parameter range, indicating that the scalar spectrum is still dominated by the standard vacuum contribution. 
As $g$ increases, however, $R_s$ is significantly enhanced. For sufficiently strong coupling, the sourced contribution becomes comparable to or even larger than the vacuum contribution. 
In addition, apart from a weak non-monotonic structure in the small-$\tilde m$ region for relatively large $g$, $R_s$ generally increases with $\tilde m$. 
This behavior reflects the enhancement of fermion production and of the associated thermal and acoustic fluctuations as the Yukawa interaction becomes stronger.

The physical mechanism of this enhancement can be interpreted from Eq. \eqref{cor_R_2}. 
The sourced scalar perturbation contains both the direct thermal-noise contribution and the thermal-acoustic contribution generated by the propagation of fluctuations through the fermion fluid. 
The latter exactly depends on the acoustic velocity $c_s^2$. Therefore, stronger Yukawa interaction enhances the scalar perturbation through two related channels: 
one is the production of fermions, which determines the strength of the thermal source, and the other is the modification of the thermodynamic and acoustic properties of the fermion fluid.
The weak oscillating feature appearing for stronger coupling is consequently related to the competition between the effective-mass and thermal effects discussed, as shown the local structure of $c_s^2$ in Fig. \ref{cs2_Fig}. 
The scalar-source terms in Eq. \eqref{cor_R_2} explicitly contain $c_s^2$ and the acoustic integrals, so this thermodynamic behavior is naturally transferred to $R_s$. 

Figure \ref{R_h_Fig} shows a similar trend of the sourced tensor spectrum. 
The ratio $R_h$ increases rapidly with both $g$ and $\tilde m$, and it is particularly pronounced in the strong-coupling and large-effective-mass regime. 
Unlike scalar perturbations, in our calculations, tensor modes are sourced by the transverse-traceless anisotropic stress of the fermion field, as derived in Appendix \ref{tensor_app}.
The correlation function contains the fermionic mode functions and Fermi-Dirac occupation factors, making the tensor source increasingly important when fermion production is enhanced and relativity of fermion fluid becomes stronger.
Thus, the thermal effect is a physical mechanism that cannot be neglected in the production of primordial gravitational waves within the warm inflationary scenario. 

The sourced scalar and tensor perturbations are enhanced at different magnitude. 
In the parameter region considered here, the relative enhancement of the total scalar spectrum is generally more efficient than that of the tensor spectrum. 
This difference arises from their different sourcing mechanisms. 
Scalar perturbations couple directly to the dissipative thermal noise and are further affected by potential acoustic propagation through the fermion fluid,
whereas tensor modes respond only to the transverse-traceless component of the fermionic anisotropic stress. 
The latter is also gravitationally suppressed by powers of the Planck mass. 
As a result, fermion production does not enhance scalar and tensor fluctuations equally, which becomes essential for the tensor-to-scalar ratio discussed below.

\subsection{Tensor-to-scalar Ratio}

\begin{figure}
	\center
	\includegraphics[width=.5\linewidth]{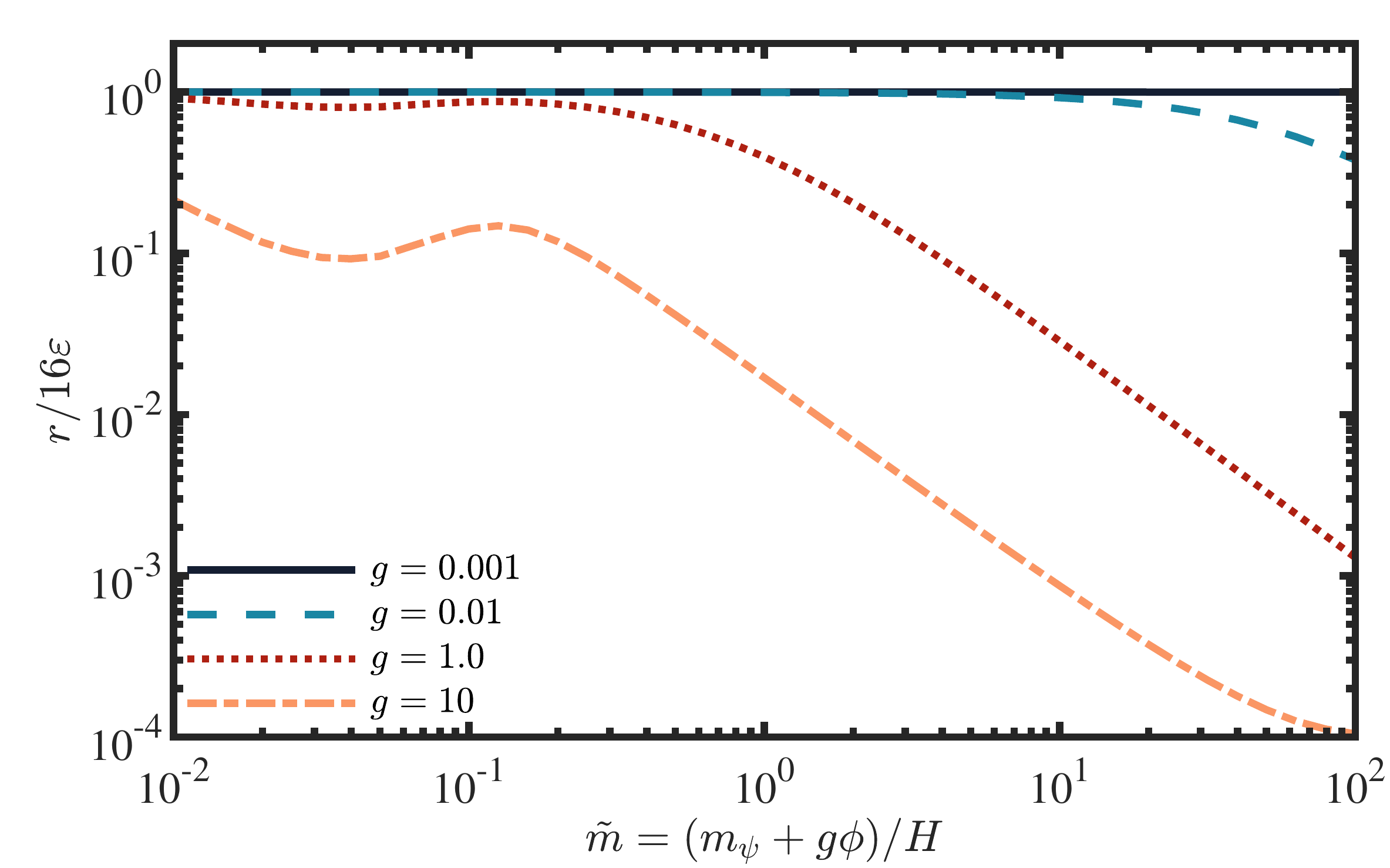}
	\caption{Tensor-to-scalar ratio $r=\mathcal P_h/\mathcal P_s$ with respect to the dimensionless effective mass $\tilde m=(m_\psi+g\phi)/H$ for various values of Yukawa interaction strength $g$. }
	\label{tensor_to_scalar_Fig}
\end{figure}

Tensor-to-scalar ratio is expressed as
\begin{align}
    r=\frac{\mathcal{P}_h^{(v)}+\mathcal{P}_h^{(s)}}{\mathcal{P}_{\mathcal{R}}^{(v)}+\mathcal{P}_{\mathcal{R}}^{(s)}}=16\varepsilon\frac{1+R_h}{1+R_s}, \label{r}
\end{align}
where $\varepsilon$ is the slow-roll parameter with definition $\varepsilon=-\dot H/H^2$. 
Figure \ref{tensor_to_scalar_Fig} presents the normalized tensor-to-scalar ratio $r/16\varepsilon$ for different values of the Yukawa coupling strength $g$. 
In the weak-coupling limit, $R_s$ and $R_h$ are both much smaller than unity, and Eq. \eqref{r} therefore reduces to the standard single-field result, i.e., $ r\simeq16\varepsilon$. 
This behavior is clearly seen for the smallest values of $g$, for which $r/16\varepsilon$ remains close to unity over a broad range of $\tilde m$.

As the Yukawa coupling becomes stronger, a substantial suppression of the tensor-to-scalar ratio appears. 
The suppression becomes more significant with increasing $\tilde m$, particularly for $g\gtrsim1$. 
According to Eq. \eqref{r}, this behavior is controlled by the competition between the sourced tensor and scalar contributions. 
Although fermion production enhances both $R_h$ and $R_s$, the sourced scalar component grows more efficiently over the relevant parameter region. 
Thus $1+R_s$ increases faster than $1+R_h$, leading to the ratio $r/16\varepsilon$ decrease.
Physically, the Yukawa interaction transfers energy from the inflaton scalar fields to the thermalized fermions deirectly, and
the produced fermionic bath generates additional scalar fluctuations through dissipative thermal noise and through acoustic propagation inside the fermion fluid. 
On the other hand, tensor perturbations are also generated by the anisotropic stress of the produced fermions, but this channel is less efficient because only the transverse-traceless component of the fermionic stress tensor can source gravitational waves. 
Therefore, increasing the efficiency of fermion production enhances the scalar spectrum more strongly than the tensor spectrum and consequently suppresses the observable tensor-to-scalar ratio.

The dependence on $\tilde m$ can be interpreted in the same way. 
Increasing the effective fermion mass changes the temperature and mode structure of the produced fermions, showing similar effects as the increasing Yukawa interaction. 
These effects increase the sourced perturbations, but their impact is stronger in the scalar sector. 
The small non-monotonic structure visible in the strong-coupling curves at relatively small $\tilde m$ is also consistent with the competing mass and thermal effects discussed for the acoustic velocity in Fig. \ref{cs2_Fig}. 
At larger $\tilde m$, the scalar enhancement dominates and tensor-to-scalar ratio decreases rapidly. 

These results reveal that thermalized fermions can change the typical relation between the slow-roll parameter and the tensor-to-scalar ratio. 
In the weak-coupling regime, the standard result $r\simeq16\varepsilon$ remains, whereas sufficiently strong Yukawa interactions may strongly reduce $r$. 
The fermion sector therefore provides a mechanism by which an inflationary background with a given value of $\varepsilon$ can produce a considerably smaller tensor-to-scalar ratio than predicted by the corresponding vacuum single-field scenario.

\section{\label{conclusion}Conclusions and outlook}    

In this work, we have investigated fermion production and its impact on primordial perturbations in a warm inflationary scenario with a Yukawa interaction between the inflaton and Dirac fermions. 
By treating the thermalized fermions as a relativistic fluid, we derived the coupled perturbation equations including dissipative effects and thermal fluctuations. 
The scalar perturbation equation was calculated using Green’s functions, which allows to identify different physical contributions arising from the interaction between thermal fluctuations and fermionic acoustic propagation.

Our analysis reveals that the sourced scalar perturbation contains three physically distinct processes: direct thermal fluctuations, thermal–acoustic mixed propagation, and higher-order scattering effects. 
The thermal–acoustic contribution is controlled by the acoustic properties of the fermion fluid and becomes significant only in the relativistic regime. 
This behavior comes from the competition between the gravitational attraction of the produced fermions and the positive degeneracy pressure induced by the Pauli exclusion principle, 
which enables the propagation of collective acoustic modes in the fermionic medium.

We further analyzed the particle density and acoustic velocity of the thermalized fermions. 
The fermion abundance is enhanced by increasing the Yukawa coupling strength, with strong coupling leading to significant deviations from the adiabatic production regime. 
The effective mass and thermal effects together determine the relativistic properties of the fermion fluid. 
In particular, their competition produces a non-monotonic structure in the acoustic velocity, reflecting the interplay between mass-induced suppression of relativistic behavior and temperature-induced enhancement.

The sourced scalar and tensor spectra were subsequently calculated. We find that stronger Yukawa interactions enhance both scalar and tensor perturbations through fermion production. 
However, scalar perturbations receive a more efficient enhancement because they are directly coupled to dissipative thermal noise and additionally affected by acoustic propagation inside the fermion fluid, 
whereas tensor perturbations are sourced only through the transverse-traceless component of the fermionic anisotropic stress. 
As a result, the tensor-to-scalar ratio can be significantly suppressed compared with the standard single-field warm inflation prediction. 
In the weak-coupling limit, our results recover the conventional relation \(r\simeq16\epsilon\), while sufficiently strong Yukawa interactions provide a mechanism for reducing $r$ even for a fixed slow-roll parameter.

The present work provides a framework for understanding how thermalized fermions modify primordial fluctuations during warm inflation. Several directions deserve further investigation. 
First, the current analysis focuses on the relativistic thermal regime, where the fermion bath can be effectively described as a perfect fluid. 
Extending the treatment beyond this approximation, including finite-mass and non-equilibrium effects, may reveal additional signatures associated with fermionic particle production.
Second, connecting the modified scalar and tensor spectra obtained here with detailed cosmic microwave background constraints and future gravitational-wave observations would provide a more comprehensive test of fermion-assisted warm inflation. Such studies may clarify whether thermalized fermions can leave distinctive signatures on primordial observables and offer new insights into the particle physics underlying the inflationary epoch. 
In addition, the effects of different fermionic species, more general Yukawa structures, and possible chemical potentials deserve further exploration.

\appendix
\section{Derivation of the dynamical equation of scalar perturbation \label{Dev_Phi}}

In this section, we derive the dynamical equation \eqref{Phi_int}. Equation \eqref{E_z} is a linear inhomogeneous equation, whose solution is given by
\begin{align}
	\mathcal{E} = c_{1} \mathrm e^{\int_{\infty}^{z} z^{-1} \mathrm d z}-z \int_{\infty}^{z} \frac{\mathrm d z_1}{z_1}\left[(1+w) H z_1 \Theta(z_1)+\frac{\xi(z_1)}{\Upsilon \dot{\phi}_{0} z_1}\right],\label{E_1}
\end{align}
where $c_1$ is the undetermined coefficient. The requirement that the perturbations remain finite demands the convergence of each term, which forces $c_1=0$.
Likewise, the solution of Eq. \eqref{Theta_z} reads
\begin{align}
	\Theta = z^{2} \int_{\infty}^{z} \frac{\mathrm d z_1}{z_1^2}\left[-\dfrac{c_{s}^{2}}{1+w} \frac{\mathcal{E}(z_1)}{H z_1}+\dfrac{3}{z_1} \Phi(z_1)\right]. \label{Theta_1}
\end{align}
Substituting Eq. \eqref{Theta_1} into Eq. \eqref{E_1} yields
\begin{align}
	\mathcal{E} &= -(1+\omega) H z \int_{\infty}^{z} \mathrm{d} z_1 z_1^2 \int_{\infty}^{z_1} \frac{\mathrm{d} z_2}{z_2^3}\left(-\frac{c_s^2}{1+w} \frac{\mathcal{E}}{H}+3 \Phi+\frac{3 \xi}{\Upsilon \dot{\phi}_0 H}\right)(z_2). \label{E_2}
\end{align}
For convenience, we introduce the following variables:
\begin{align}
	\Psi = \mathcal{E} / H,\quad f(z)=\int_{\infty}^{z} \mathrm{d} z_{1} z_{1}^{2} \int_{\infty}^{z_{1}} \mathrm{d} z_{2} z_{2}^{-3} \Phi\left(z_{2}\right),
    \quad g(z) = \int_{\infty}^{z} \mathrm{d} z_{1} z_{1}^{2} \int_{\infty}^{z_{1}} \mathrm{d} z_{2} z_{2}^{-3} \xi\left(z_{2}\right).  \label{def}
\end{align}
With these definitions, the dynamical equation \eqref{E_2} simplifies to
\begin{align}
	\Psi = -c_s^2 z \int_0^z \mathrm{d}z_1 \, z_1^2 \int_{\infty}^{z_1} \frac{\mathrm{d}z_2}{z_2^3} \Psi - 3(1+w)z f(z) - \frac{3z g(z)}{\Upsilon \dot{\phi}_{0} H}. \label{Psi_1}
\end{align}
Multiplying $1/z$ of Eq. \eqref{Psi_1} and differentiating with respect to $z$ to eliminate the integral of $z_1$, it arrives
\begin{align}	
	-\frac{\Psi}{z^2}+\frac{\Psi'}{z} = -c_s^2 z^2 \int_{\infty}^{z} \dfrac{\mathrm{d} z_2}{z_2^3}\Psi-3(1+w) f'-\dfrac{3 g'}{\Upsilon \dot{\phi}_{0} H}, \label{Psi_2}
\end{align}
where terms involving slow-roll parameters (in particular those arising from derivatives of the Hubble parameter $H$) have been neglected.
Furthermore, multiplying Eq. \eqref{Psi_2} by $1/z^2$ and differentiating with respect to $z$, and subsequently multiplying both sides by $z^4$, we arrive at 
\begin{align}
	\Psi''-\dfrac{4}{z} \Psi'+\left(c_s^2+\dfrac{4}{z^2}\right) \Psi = -3(1+w)(z f''-2 f')-\dfrac{3}{\Upsilon \dot{\phi}_0 H}(z g''-2 g'). \label{Psi_3}
\end{align}
From the definitions of $f$ and $g$ it follows at once that $z f''-2f'=\Phi$ and $z g''-2g'=\xi$.
Consequently, the equation of motion for $\Psi$ reduces to
\begin{align}
	\Psi''-\dfrac{4}{z} \Psi'+\left(c_s^2+\dfrac{4}{z^2}\right) \Psi = -3(1+w)\Phi-\dfrac{3}{\Upsilon \dot{\phi}_0 H}\xi. \label{Psi_4}
\end{align}
This is again a linear inhomogeneous equation. 

The dynamical equation \eqref{Psi_4} can be solved by means of the Green’s function method. 
The linear differential equation 
\begin{align}
	y''+\dfrac{1-2 \alpha}{x} y'+\left(\beta^{2}+\dfrac{\alpha^{2}-\mu^{2}}{x^{2}}\right) y = 0
\end{align}
possesses two linearly independent solutions of the form $y = x^\alpha Z_\mu (\beta x)$, where $Z_\mu (\beta x)$ denotes an arbitrary Bessel function. 
The parameters appearing in this equation correspond to those of Eq. \eqref{Psi_4} with $\beta = c_s$, $\alpha = \frac{5}{2}$ and $\mu = \frac{3}{2}$, respectively. 
Therefore, the homogeneous part of Eq. \eqref{Psi_4} admits the two linearly independent solutions
\begin{align}
	\Phi_{\text{lin}} = (c_s z)^{5/2}\mathrm J_{3/2}(c_s z),\quad  (c_s z)^{5/2}\mathrm Y_{3/2}(c_s z),  \label{Psi_lin}
\end{align}
where $\mathrm J_{\nu}(z)$ is the Bessel function of the first kind of order $\nu$ and $\mathrm Y_{\nu}(z)$ is the Neumann function of order $\nu$. 
The Bessel-function order $\nu=3/2$ is closely connected with the spatial propagation of acoustic waves in de Sitter spacetime. 
Applying the Green's function method, the solution of equation \eqref{Psi_4} may be written in a form of integral:
\begin{align}
	\Psi = -\int\mathrm d z_1 \, G_{c_s, k}^{(\Psi)}(z, z_1) \left[ 3(1+\omega) \Phi(z_1) + \frac{\xi(z_1)}{\gamma \dot{\phi}_{0} H} \right], \label{Psi_sol}
\end{align}
with the Green’s function given by 
\begin{align}
	G_{c_s,k}^{(\Psi)}(z, z_1) = \frac{z}{c_s z_1^3} \left[ (1 + c_s^2 z z_1) \sin c_s (z_1 - z) + c_s (z - z_1) \cos  (z_1 - z) \right] \theta(z_1 - z),  \label{G_Psi}
\end{align}
where $\theta(z)$ denotes the Heaviside step function. 
The deviations of Green's function are explicitly presented in Ref. \cite{Green1}.

Now, we turn to solve the dynamical equation about $\Phi$, expressed in Eq. \eqref{Phi_z}. 
Analogously, the Green’s function yields the integral representation 
\begin{align}
	\Phi(z) = \int \frac{\mathrm d z_1}{z_1^2} G_k^{(\Phi)}(z, z_1) \left[ -\frac{3Q\Psi(z_1)}{(1+\omega)} + \frac{\xi(z_1)}{\dot{\phi}_{0} H^2} \right]. \label{Phi_1}
\end{align}
The Green’s function $G_k^{(\Phi)}(z, z_1)$ is given in Refs. \cite{Green1,Green2} by
\begin{align}
	G_k^{(\Phi)}(z, z_1) = \frac{z^\nu z_1^\nu}{z_1^{2\nu} (2/\pi z_1)} \left[\mathrm J_\nu(z) \mathrm Y_\nu(z_1) - \mathrm J_\nu(z_1) \mathrm Y_\nu(z) \right] \theta(z_1 - z), \label{G_Phi}
\end{align}
where the order of the Bessel functions reads 
\begin{align}
		\nu = \frac{Q}{2} + \frac{3}{2}.
\end{align}
Substitution of the expression for $\Psi$ obtained in Eq. \eqref{Psi_sol} into Eq. \eqref{Phi_1} finally reduces the equation of motion to a Volterra integral equation of the second kind, which is precisely Eq. \eqref{Phi_int}.

\section{Expressions of fermion density and acoustic velocity \label{Expre} }

For the Dirac fields in a quasi-de Sitter spacetime (within the slow-roll approximation) described by the Lagrangian \eqref{L}, the exact solutions for the different chiral and spin states are given by
\begin{subequations} \label{sol_psi_J}
\begin{align}
    &u_{s=+}^{h=+} =  \Bigg(\frac{-\pi k\tau}{2\cosh{\pi\tilde{\xi}}}\Bigg)^{1/2}
        \begin{pmatrix}\begin{pmatrix}1\\0\end{pmatrix}\mathrm J_{-\frac12+\tilde{\eta}+\mathrm{i}\tilde{\xi}}(-k\tau)\\[1mm]
        -\frac{\mathrm{i}}k\begin{pmatrix}{k_z}\\{k_+}\end{pmatrix}\mathrm J_{\frac12+\tilde{\eta}+\mathrm{i}\tilde{\xi}}(-k\tau)\end{pmatrix}, \\
    &u_{s=-}^{h=+} =  \Bigg(\frac{-\pi k\tau}{2\cosh{\pi\tilde{\xi}}}\Bigg)^{1/2}
        \begin{pmatrix}\begin{pmatrix}0\\1\end{pmatrix}\mathrm J_{-\frac12+\tilde{\eta}+\mathrm{i}\tilde{\xi}}(-k\tau)\\[1mm]
     -\frac{\mathrm{i}}{k}\begin{pmatrix}{k_-}\\{-k_z}\end{pmatrix}\mathrm J_{\frac12+\tilde{\eta}+\mathrm{i}\tilde{\xi}}(-k\tau)\end{pmatrix},  \\
    &u_{s=+}^{h=-} =  \Bigg(\frac{-\pi k\tau}{2\cosh{\pi\tilde{\xi}}}\Bigg)^{1/2}
        \begin{pmatrix}\begin{pmatrix}1\\0\end{pmatrix}\mathrm J_{\frac12-\tilde{\eta}-\mathrm{i}\tilde{\xi}}(-k\tau)\\[1mm]
    \frac{\mathrm{i}}{k}\begin{pmatrix}{k_z}\\{k_+}\end{pmatrix}\mathrm J_{-\frac12-\tilde{\eta}-\mathrm{i}\tilde{\xi}}(-k\tau)\end{pmatrix},  \\
    &u_{s=-}^{h=-} =  \Bigg(\frac{-\pi k\tau}{2\cosh{\pi\tilde{\xi}}}\Bigg)^{1/2}
        \begin{pmatrix}\begin{pmatrix}0\\1\end{pmatrix}\mathrm J_{\frac12-\tilde{\eta}-\mathrm{i}\tilde{\xi}}(-k\tau)\\[1mm]
    \frac{\mathrm{i}}{k}\begin{pmatrix}{k_-}\\{-k_z}\end{pmatrix}\mathrm J_{-\frac12-\tilde{\eta}-\mathrm{i}\tilde{\xi}}(-k\tau)\end{pmatrix}, 
\end{align}
\end{subequations}
where $h=\pm$ label the right- and left-handed states, and $s=\pm$ label the spin-up and spin-down states. Here $\mathrm J_\nu(z)$ remains the Bessel function of the first kind. 
The parameters appearing in Eqs. \eqref{sol_psi_J} are defined by
\begin{align}
    \tilde\eta=\frac{\tilde\delta\tilde m}{\frac14+\tilde m^2}, \quad
    \tilde\xi=\tilde m-\frac{\tilde\delta}{\frac12+2\tilde m^2}, \quad
    \tilde{m}=\frac{m+g\phi}{H},\quad 2\tilde{\delta}=\frac{g\dot{\phi}+\dot{H}\tilde m}{H^2}. \label{para}
\end{align}
Detailed derivations of these solutions can be found in Ref. \cite{Dirac4}. 

The quantized Dirac field at finite time admits the mode expansion
\begin{align}
	\hat{\psi}(\mathbf{x}, \tau) &= \sum_{s=\pm} \int \frac{\textrm{d}^3{k}}{(2\pi a)^{3/2}} \textrm{e}^{\textrm{i}\mathbf{k}\cdot\mathbf{x}} 
        \left[ \hat{b}_s(\mathbf{k}) u_s^{h=+}(\mathbf{k},\tau) + \hat{d}_s^\dagger(-\mathbf{k}) u_s^{h=-}(-\mathbf{k},\tau) \right].  \label{qunta_D}
\end{align}
Here $\hat{b}_s(\mathbf{k})$ and $\hat{d}_s^\dagger(\mathbf{k})$ denote the annihilation and creation operators for particles and antiparticles, respectively, carrying spin $s$ and momentum $\mathbf{k}$. 
Their anticommutation relations are given by
\begin{align}
    \left\{\hat{b}_s(\mathbf{k}),\hat{b}_{s^{\prime}}^\dagger(\mathbf{k}^{\prime})\right\}=\left\{\hat{d}_s(\mathbf{k}),\hat{d}_{s^{\prime}}^\dagger(\mathbf{k}^{\prime})\right\} 
        =(2\pi)^3\delta_{ss^{\prime}}\delta^3(\mathbf{k}-\mathbf{k}^{\prime}). \label{commu}
\end{align}

\subsection{Net particle density of fermion \label{n_psi_app}}

The net density of fermion pairs in comoving coordinates is defined as the ensemble average of the Dirac field and its Hermitian conjugate,
\begin{align}
	n_\psi(x) &=  \langle \hat{\psi}^\dagger(x) \hat{\psi}(x) \rangle  \nonumber \\
	&= \frac{\pi}{2a^3 \cosh \pi \tilde{\xi}} \int \frac{\textrm{d}^3k}{(2\pi)^3} (-k\tau) \sum_{s,s'} 
        \left[ \langle \hat{b}_s^\dagger(\mathbf{k}) \hat{b}_{s'}(\mathbf{k}) \rangle u_s^{h=+\, \dagger}(\mathbf{k},\tau) u_{s}^{h=+}(\mathbf{k},\tau) \right.\nonumber \\
	&\qquad\qquad\qquad\qquad\qquad\qquad\qquad\qquad \left. + \langle \hat{d}_s(\mathbf{k}) \hat{d}_{s'}^\dagger(\mathbf{k}) \rangle u_s^{h=-\,\dagger}(\mathbf{k},\tau) u_{s}^{h=-}(\mathbf{k},\tau) \right]\nonumber \\
	&= \frac{4\pi^2 H^3}{\cosh \pi \tilde{\xi}} \int_0^\infty \textrm{d}(-k\tau) (-k\tau)^3 \left\{ \left[ \left| \textrm{J}_{\frac{1}{2}+\tilde{\eta} + \textrm{i}\tilde{\xi}}(-k\tau) \right|^2 + 
        \left| \textrm{J}_{-\frac{1}{2} +\tilde{\eta}+ \textrm{i}\tilde{\xi}}(-k\tau) \right|^2 \right] \langle \hat{b}_s^\dagger(\mathbf{k}) \hat{b}_{s'}(\mathbf{k}) \rangle \right. \nonumber\\
	&\qquad\qquad\qquad\qquad\qquad \left.- \left[ \left| \textrm{J}_{-\frac{1}{2}-\tilde{\eta} - \textrm{i}\tilde{\xi}}(-k\tau) \right|^2 + \left| \textrm{J}_{\frac{1}{2}-\tilde{\eta} - \textrm{i}\tilde{\xi}}(-k\tau) \right|^2 \right] 
        \langle \hat{d}_s^\dagger(\mathbf{k}) \hat{d}_{s'}(\mathbf{k}) \rangle \right\} \nonumber\\
	&\qquad\qquad\qquad + \frac{2}{a^3} \int \textrm{d}^3k \delta^3(\mathbf{0}) u_{-}^{-\,\dagger}(\mathbf{k},\tau) u_{-}^{-}(\mathbf{k},\tau),   \label{n_psi_1}
\end{align}
where $z=-k\tau$ and $\langle\cdots\rangle$ denotes the ensemble average. 
In the second equality the orthogonality relation $\tilde{u}_s^{h\,\dagger}\tilde{u}_r^{h}\propto\delta_{sr}$ has been used.
In the last equality the anticommutation relations of the operators $\hat{d}$ and $\hat{d}^\dagger$ have been employed.
Terms containing $\hat{b}^\dagger\hat{d}$ and $\hat{b}\hat{d}^\dagger$ have been omitted because they vanish upon taking the ensemble average.
The final term in the last line is an infinite constant. It corresponds to the sum of the zero-point energies of the infinite set of eigenfunctions, each contributing a factor $\tilde{u}_s^{-\,\dagger}(\mathbf{k},\tau)\tilde{u}_s^{-}(\mathbf{k},\tau)$.
Such a divergent contribution is ubiquitous in quantum field theory (analogous to the zero-point energy of the harmonic oscillator) and will be handled consistently in the subsequent calculations.
Because the term is a pure constant, it can be removed by a simple shift of the energy origin.
The physically relevant issue resides in the first (negative) term. Its presence indicates that the energy can be lowered by the creation of additional particles in the modes $u_s^{h=-}$.
The same observation will be used later when calculating the energy density and pressure.

The ensemble average of fermion operators gives 
\begin{subequations}
\begin{align}
	&\langle \hat{b}_s^\dagger(\mathbf{k}) \hat{b}_{s'}(\mathbf{k}') \rangle=  (2\pi)^3 \delta_{ss'} \delta^3(\mathbf{k}-\mathbf{k}') f(E, \mu) \\
	&\langle \hat{d}_s^\dagger(\mathbf{k}) \hat{d}_{s'}(\mathbf{k}') \rangle = (2\pi)^3 \delta_{ss'} \delta^3(\mathbf{k}-\mathbf{k}') f(E, -\mu), 
\end{align} \end{subequations}
where the Fermi–Dirac distribution $f(E,\mu)$ and the relativistic energy $E$ are defined by 
\begin{align}
	f(E, \mu) = \left[\mathrm e^{(E-\mu)/T} + 1\right]^{-1},\quad E = \sqrt{({k}/a)^2 + (m_\psi + g\phi)^2}.  \label{f_E}
\end{align}	
Expanding the Bessel functions to first order in the slow-roll parameter $\tilde{\eta}$ then yields 
\begin{align}  \label{n_psi_2}
	n_\psi = \frac{4\pi^2H^3}{\cosh \pi \tilde{m}} \int_0^\infty \textrm{d}z \, z^3 & \Bigg\{ [f(E, \mu) - f(E, -\mu)] 
        \left[ \left|\textrm{J}_{\frac{1}{2}+\textrm{i}\tilde{m}}(z)\right|^2 + \left|\textrm{J}_{-\frac{1}{2}-\textrm{i}\tilde{m}}(z)\right|^2 \right]\nonumber \\
	& + \tilde{\eta} [f(E, \mu) + f(E, -\mu)] \left[ \left. \frac{\partial \textrm{J}_\nu(z)}{\partial \nu} \right|_{\nu = \frac{1}{2}+\textrm{i}\tilde{\xi}} \textrm{J}_{\frac{1}{2}-\textrm{i}\tilde{\xi}}(z) + \text{c.c.} \right] \nonumber \\
	&\qquad\qquad\qquad\qquad + \left.\left. \frac{\partial \textrm{J}_\nu(z)}{\partial \nu} \right|_{\nu = -\frac{1}{2}+\textrm{i}\tilde{\xi}} \textrm{J}_{-\frac{1}{2}-\textrm{i}\tilde{\xi}}(z) + \text{c.c.} \right]\Bigg\}  
    + \mathcal O(\tilde{\eta}^2),
\end{align}
with $z=-k\tau$ and “c.c.” denoting the complex conjugate of the preceding term. 
The divergent contribution proportional to $\delta^3(\mathbf{0})$ has been neglected for the reasons explained above.
We now recall the series representation of the Bessel function and of its derivative with respect to the order \cite{NIST},
\begin{subequations}
\begin{align}
    \textrm{J}_\nu(z) & =\sum_{k=0}^\infty\frac{(-1)^k}{k!\Gamma(k+\nu+1)}\left(\frac z2\right)^{2k+\nu}, \\
    \frac{\partial\textrm{J}_\nu(z)}{\partial\nu} & =\textrm{J}_\nu(z)\ln(z/2)-\sum_{k=0}^\infty\frac{(-1)^k\psi_M(k+\nu+1)}{k!\Gamma(k+\nu+1)}\left(\frac z2\right)^{2k+\nu},
\end{align}\end{subequations}
where $\psi_M(z)$ is digamma function. Substitution of these expansions produces the final expression for the fermion density, 
\begin{align}
	n_\psi &= n_0 + \frac{4\pi^2 \tilde{\eta}H^3}{\cosh \pi \tilde{\xi}} \int_0^\infty \textrm{d}z \, z^3 \ln( z/2) [f(E, \mu) - f(E, -\mu)]\cdot 
        \left[ \left|\textrm{J}_{\frac{1}{2}+\textrm{i}\tilde{\xi}}(z)\right|^2 + \left|\textrm{J}_{-\frac{1}{2}-\textrm{i}\tilde{\xi}}(z)\right|^2 \right]\nonumber \\
	&\quad - \frac{4\pi^2 \tilde{\eta}H^3}{\cosh \pi \tilde{\xi}} \text{Re} \sum_{h=\pm} \sum_{k=0}^\infty \sum_{l=0}^\infty \frac{(-1)^{k+l} \psi_M (h/2 + \textrm{i}\tilde{\xi} + k + 1) 
        \int_0^\infty \textrm{d}z \, z^{2k+2\ell+h+3} [f(E, \mu) - f(E, -\mu)]}{k!l!2^{2k+2l+h-1} \Gamma(h/2 + \textrm{i}\tilde{\xi} + k + 1) \Gamma(h/2 - \textrm{i}\tilde{\xi} + l + 1)}  \nonumber \\
    &\quad + \mathcal O(\tilde{\eta}^2).  \label{n_psi_3}
\end{align}
Here, $n_0$ is the net particle density evaluated at $\tilde{\eta}=0$,
\begin{align}
    n_0= &\frac{4\pi^2H^3}{\cosh \pi \tilde{m}} \int_0^\infty \textrm{d}z \, z^3  \left[ \left| \textrm{J}_{\frac{1}{2}+\textrm{i}\tilde{m}}(z) \right|^2 + \left| \textrm{J}_{-\frac{1}{2}-\textrm{i}\tilde{m}}(z) \right|^2 \right] \cdot \nonumber \\
        &\qquad\qquad\qquad\left[ \frac{1}{\textrm{e}^{\left(\sqrt{{z}^2 + \tilde{m}^2}-\tilde m\right)H/T} + 1}-\frac{1}{\textrm{e}^{\left(\sqrt{{z}^2 + \tilde{m}^2}+\tilde m\right)H/T} + 1}  \right],  \label{n0_app}
\end{align}
where the chemical potential is the effective mass for the relativistic limit, i.e., $\mu=m_\psi+g\phi$.
This physical variable corresponds to the leading-order value in exact de Sitter spacetime. 

\subsection{Energy density and pressure of fermion \label{rho_p_app}}

The Hamiltonian of Dirac field is expressed as
\begin{align}
    \hat H &= \int \mathrm d^3x\, \mathcal{H} = 
    \int \mathrm d^3x\left[\mathrm i\hat \psi^\dagger(x)\partial_0\hat \psi(x)-\hat{\bar{\psi}}(x)(\mathrm i\bar \gamma^\mu\partial_\mu-m_\psi-g\phi)\hat \psi(x)\right]  \nonumber\\ 
    &=\sum_{s=\pm}\int\frac{\mathrm d^3k}{(2\pi a)^3} \sqrt{(k/a)^2+(m_\psi+g\phi)^2}
    \left[\hat{b}^\dagger_s(\mathbf{k})\hat{b}_s(\mathbf{k})\tilde{u}_s^{h=+\ \dagger}(\mathbf{k},\tau)\tilde{u}_s^{h=+}(\mathbf{k},\tau) \right.  \nonumber\\ 
    &\qquad\qquad\qquad\qquad\qquad\qquad\qquad\qquad\qquad\quad \left. +\hat{d}^\dagger_s(\mathbf{k})\hat{d}_s(\mathbf{k})\tilde{u}_s^{h=-\ \dagger}(\mathbf{k},\tau)\tilde{u}_s^{h=-}(\mathbf{k},\tau) \right]  \nonumber\\ 
    &\qquad\qquad\qquad\qquad\qquad -\frac{2}{a^3}\int\mathrm d^3k\, \delta^3(\mathbf 0)\sqrt{(k/a)^2+(m_\psi+g\phi)^2}\ \tilde{u}_-^{-\ \dagger}(\mathbf{k},\tau)\tilde{u}_-^{-}(\mathbf{k},\tau).
\end{align}
The last line of the expression above also contains a term proportional to $\delta(\mathbf{0})$. As discussed previously, this contribution may be identified with the zero-point energy and is therefore discarded.
In the derivation, we retain only the leading-order terms by working in exact de Sitter spacetime, where $\tilde\eta=0$ and $\tilde\xi=\tilde m$.
The energy density of the Dirac field is the vacuum expectation value of the Hamiltonian density, given by
\begin{align}
	\rho_\psi &= \langle \mathcal{H} \rangle = \frac{4\pi^2 H^4}{\cosh \pi \tilde{m}} \int_0^\infty \textrm{d}z {z^3}{\sqrt{z^2 + \tilde{m}^2}} 
        \left[ \left| \textrm{J}_{\frac{1}{2} + \textrm{i}\tilde{m}}(z) \right|^2 + \left| \textrm{J}_{-\frac{1}{2} - \textrm{i}\tilde{m}}(z) \right|^2 \right] \left[ f(E, \mu) + f(E, -\mu) \right]. \label{rho_app} 
\end{align}
Similarly, the pressure of the Dirac field reads
\begin{align}
	p_\psi &= \frac{\pi}{a^5} \frac{1}{\cosh \pi \tilde{m}} \int {\textrm{d}^3{k}} \frac{(-k\tau) {k}^2}{3\sqrt{({k}/a)^2 + (m_\psi + g\phi)^2}} 
        \left[ \left| \textrm{J}_{\frac{1}{2} + \textrm{i}\tilde{m}}(-k\tau) \right|^2 + \left| \textrm{J}_{-\frac{1}{2} + \textrm{i}\tilde{m}}(z) \right|^2 \right]\left[ f(E, \mu) + f(E, -\mu) \right]  \nonumber\\ 
	&= \frac{4\pi^2 H^4}{\cosh \pi \tilde{m}} \int_0^\infty \textrm{d}z \frac{z^5}{3\sqrt{z^2 + \tilde{m}^2}} \left[ \left| \textrm{J}_{\frac{1}{2} + \textrm{i}\tilde{m}}(z) \right|^2 + 
        \left| \textrm{J}_{-\frac{1}{2} - \textrm{i}\tilde{m}}(z) \right|^2 \right] \left[ f(E, \mu) + f(E, -\mu) \right], \label{p_psi_app} 
\end{align}
where the divergent $\delta(\mathbf{0})$ contribution has again been omitted.
With these expressions, the acoustic velocity of the ideal fermion fluid is obtained as $c_s^2=p_\psi/\rho_\psi$.

\section{Expression of the tensor perturbation correlator reduced from Dirac fields \label{tensor_app}}

The equation of motion for tensor perturbations follows from the Einstein equations and takes the form
\begin{align}
    \left(\frac{\partial^2}{\partial\tau^2}+k^2-\frac{\mathrm d^2a/\mathrm d\tau^2}{a}\right)(ah_{ij})=\frac{2a^3}{M_p^2}\Pi_{ij,mn}T^{mn}(\mathbf{k},\tau),
\end{align}
where \(M_p=(8\pi G)^{-1/2}\) is the Planck mass and \(\Pi_{ij,mn}(\mathbf{k})=P_{im}(\mathbf{k})P_{jn}(\mathbf{k})-\frac12 P_{ij}(\mathbf{k})P_{mn}(\mathbf{k})\) 
is the transverse-traceless projector constructed from the spatial projector $P_{ij}(\mathbf{k})=\delta_{ij}-k_i k_j/k^2$. 
The energy-momentum tensor of the fermionic field is obtained by varying the Dirac Lagrangian \eqref{L_D} with respect to the metric $g_{mn}$ \cite{Dirac1,Dirac2}:
\begin{align}
    T^{mn}(\mathbf{x},\tau)&=\frac{2}{\sqrt{-g}}\frac{\delta\mathcal{L}}{\delta g_{mn}}\nonumber\\
    &=\frac{\mathrm i}{4a}\bar\psi\bigl[\gamma^m\overrightarrow{D}^n+\overleftarrow{D}^n\gamma^m\bigr]\psi+\frac12 g^{mn}\mathcal{L} + (m\leftrightarrow n),
\label{T_x}
\end{align}
where $\bar\psi\overleftarrow{D}^n=-(\partial_\mu\bar\psi-\bar\psi\Gamma_\mu)$.
After a Fourier transformation the energy-momentum tensor in momentum space becomes
\begin{align}
    T_{mn}(\mathbf{k},\tau)= &\frac{\mathrm i}{4a}\int{\mathrm{d}^3p}\, \bar{\psi}(\mathbf{k}-\mathbf{p},\tau) 
        \left[\gamma^mp^n-\gamma^n(k-p)^m\right]\psi(\mathbf{p})\nonumber\\
    &+\frac{\mathrm i }{4}\frac{\dot{a}}{a}\int{\mathrm{d}^3p}\, \bar{\psi}(\mathbf{k}-\mathbf{p},\tau)[\gamma^m\gamma^0\gamma^n+\gamma^0\gamma^n\gamma^m]\psi(\mathbf{p}) \nonumber\\
    &+(m\leftrightarrow n). \label{T_k}
\end{align}
The term proportional to $g^{mn}\mathcal{L}$ in Eq.~\eqref{T_x} has been dropped, since its contraction with the projector vanishes identically under the traceless condition $\Pi_{ij,mn}g^{mn}=0$.

We decompose the tensor perturbation into polarized modes
\begin{align}
h_{ij}(\mathbf{k},\tau)=h_+(\mathbf{k},\tau)\,e_{ij}^+(\mathbf{k})+h_-(\mathbf{k},\tau)\,e_{ij}^-(\mathbf{k}),
\end{align}
where the polarization tensors \(e_{ij}^\pm\) satisfy $e_{ij}^\lambda(\mathbf{k})\,e_{ij}^{\lambda'}(\mathbf{k})^\dagger=\delta_{\lambda\lambda'}$, 
$e_{ij}^\lambda(\mathbf{k})k^i=0$, and $e_{ij}^\pm(\hat{\mathbf{k}})\Pi_{mn}^{ij}(\mathbf{k})=e_{mn}^\pm(\mathbf{k})$.
They are constructed from the polarization vectors via $e_{ij}^\pm=\frac1{\sqrt2}\,\epsilon_i^{(\pm)}\otimes\epsilon_j^{(\pm)}$, with $\epsilon_i^\pm=(1,\pm \mathrm i,0)^T$. 
Employing the Green’s-function method, the solution for the tensor-mode operator \(\hat h_\lambda\) reduced from Dirac fields is
\begin{align}
    \hat{h}^{(s)}_\lambda(\mathbf{k},\tau) = & \frac2{M_p^2 a(\tau)}e_{mn}^\lambda(\mathbf{k})^\dagger \int\mathrm{d}\tau^{\prime}G_{\mathbf{k}}(\tau,\tau^{\prime})a^2(\tau^{\prime})
        \int\operatorname{d}^3p\, \frac{\mathrm i}{4}\hat{\bar{\psi}}(\mathbf{k}-\mathbf{p},\tau^{\prime}) \cdot \nonumber\\
    &\left[\gamma^mp^n-\gamma^n(k-p)^m+\frac{\dot{a}}{2}(\gamma^m\gamma^0\gamma^n+\gamma^0\gamma^n\gamma^m)\right]\hat{\psi}(\mathbf{p},\tau') \nonumber\\
    &+(m\leftrightarrow n),   \label{h}
\end{align}
with the Green’s function \cite{Green1,Green2}
\begin{align}
    G_k(\tau,\tau^{\prime})=\frac{1}{k^3\tau\tau^{\prime}}\left[(1+k^2\tau\tau^{\prime})\sin k(\tau-\tau^{\prime})+k(\tau^{\prime}-\tau)\cos k(\tau-\tau^{\prime})\right]\theta(\tau-\tau^{\prime}).
\end{align}
Thus, the two-point correlation function of the sourced tensor perturbation is 
\begin{align}
	\left\langle \hat{h}_{\lambda}^{(s)}(\mathbf{k}, \tau) \, \hat{h}_{\lambda}^{(s)}(\mathbf{k}', \tau)^\dagger \right\rangle 
	   =& \frac{4}{M_p^4 a^2(\tau)} \int \mathrm{d}\tau_1 \, G_k(\tau, \tau_1) a^{2}(\tau_1) \int \mathrm{d}\tau_2 \, G_{k'}(\tau, \tau_2) a^{2}(\tau_2) \cdot \nonumber \\
	&\int \frac{\mathrm{d}^3p}{(2\pi)^3} \int \frac{\mathrm{d}^3p'}{(2\pi)^3} \left\langle \hat{\bar{\psi}}(\mathbf{k}-\mathbf{p}) e_{ij}^{\lambda}(\hat{\mathbf k}) \gamma^i p^j  \hat{\psi}(\mathbf{p})
		\left[ \hat{\bar{\psi}}(\mathbf{k}'-\mathbf{p}') e_{mn}^{\lambda}(\hat{\mathbf k}') \gamma^m p'^n  \hat{\psi}(\mathbf{p}') \right]^\dagger \right\rangle. 
\end{align}
Noticing that $(\gamma^m)^\dagger=\gamma^0\gamma^m\gamma^0$, the ensemble average in the above equation then becomes 
\begin{align}
    &\left\langle \hat{\bar{\psi}}(\mathbf{k}-\mathbf{p})e_{ij}^\lambda(\hat{k})\gamma^i p^j\hat{\psi}(\mathbf{p})
		\left[ \hat{\bar{\psi}}(\mathbf{k}'-\mathbf{p}')e_{mn}^\lambda(\hat{k}')\gamma^m p'^n\hat{\psi}(\mathbf{p}') \right]^\dagger \right\rangle \nonumber \\
	&= e_{ij}^{\lambda}(\hat{\mathbf k}) e_{mn}^{\bar{\lambda}}(\hat{\mathbf{k}}')
		\left\langle \hat{\bar{\psi}}(\mathbf{k}-\mathbf{p}) \gamma^i p^j \hat{\psi}(\mathbf{p})\hat{\bar{\psi}}(\mathbf{p}')
		\gamma^m p'^n \hat{\psi}(\mathbf{k}'-\mathbf{p}') \right\rangle \nonumber \\
	&= e_{ij}^{\lambda}(\hat{k}) e_{mn}^{\bar{\lambda}}(\hat{k}')\Big[ \left\langle \hat{\bar{\psi}}(\mathbf{k}-\mathbf{p}) \gamma^i p^j \hat{\psi}(\mathbf{p}) \right\rangle
		\left\langle \hat{\bar{\psi}}(\mathbf{p}') \gamma^m p'^n \hat{\psi}(\mathbf{k}'-\mathbf{p}') \right\rangle  \nonumber\\
    &\qquad\qquad\qquad\qquad - \left\langle \hat{\bar{\psi}}(\mathbf{k}-\mathbf{p}) \hat{\bar{\psi}}(\mathbf{p}')\right\rangle
		\left\langle \gamma^i p^j \hat{\psi}(\mathbf{p}) \gamma^m p'^n \hat{\psi}(\mathbf{k}'-\mathbf{p}') \right\rangle \nonumber \\
	&\qquad\qquad\qquad\qquad + \left\langle \hat{\bar{\psi}}(\mathbf{k}-\mathbf{p}) \gamma^m p'^n \hat{\psi}(\mathbf{k}'-\mathbf{p}') \right\rangle
		\left\langle \hat{\bar{\psi}}(\mathbf{p}') \gamma^i p^j \hat{\psi}(\mathbf{p}) \right\rangle \Big].
\end{align}
In the second equality, the Wick's theorem is applied \cite{Wick} and the commutation of two Dirac operator contributes a minus sign. 

The contraction of a polarization tensor with the gamma matrices yields
\begin{align}
    e_{ij}^\pm(\hat{\mathbf{k}})\gamma^i p^j=e_{ij}^\pm(\hat{\mathbf{k}})\gamma^j p^i=\frac{p^\pm\gamma^\pm}{\sqrt2},
\end{align}
with \(\gamma^\pm=\gamma^1\pm \mathrm i\gamma^2\) and \(p^\pm=p^1\pm \mathrm i p^2\). The vacuum expectation value of the corresponding contracted polarization term, evaluated with the fermion mode functions \eqref{sol_psi_J} and the quantized fermion field \eqref{qunta_D}, is then given by
\begin{align}
    \left\langle \bar{\psi}(\mathbf{p}',\tau_1) e_{ij}^{\pm}(\hat{k}) \gamma^i p^j \hat{\psi}(\mathbf{p},\tau_2) \right\rangle
		&= \frac{\mathrm i\pi}{2\sqrt{2}} \frac{p_{\pm}^2}{p} \frac{\delta^3(\mathbf{p}-\mathbf{p}')}{[a(\tau_1)a(\tau_2)]^{3/2}}
        \frac{(p^2 \tau_1 \tau_2)^{1/2}}{\cosh\pi \tilde{m}} \left[ f(E_p, \mu) + f(E_p, -\mu) \right] \nonumber \\
	&\quad  \left[\mathrm J_{\frac12-\mathrm{i}\tilde{m}}(-p\tau_1)\mathrm J_{-\frac12+\mathrm{i}\tilde{m}}(-p\tau_2)
        -\mathrm J_{-\frac12-\mathrm{i}\tilde{m}}(-p\tau_1)\mathrm J_{\frac12+\mathrm{i}\tilde{m}}(-p\tau_2) \right]. 
\end{align}
The Fermi-Dirac distribution $f(E_p, \mu)$ follows the expression in Eq. \eqref{f_E}. 
In addition, the contribution from the remaining gamma matrices vanishes $\gamma^m\gamma^0\gamma^n+\gamma^0\gamma^n\gamma^m+(m\leftrightarrow n)=0$ because $\{\gamma^m,\gamma^0\}=0$,
leaving us with the correlation function for the sourced tensor perturbation at large-scale limit $-k\tau\to0$:
\begin{align}
    &\left\langle \hat{h}_{\lambda}^{(s)}(\boldsymbol{k}, 0) \hat{h}_{\lambda}^{(s)}(\boldsymbol{k}', 0)^\dagger \right\rangle
		= \left\langle \hat{h}_{\lambda}^{(s)}(\boldsymbol{k}, \tau) \hat{h}_{\lambda}^{(s)}(\boldsymbol{k}', \tau)^\dagger \right\rangle_{{k}, {k}' \ll a_* H_*} \nonumber \\
	&= \frac{\pi^2 }{2 M_P^4 a^2(\tau)}  \frac{1}{\cosh^2\pi \tilde{m}}  \delta^3(\boldsymbol{k} - \boldsymbol{k}')
		\int \mathrm{d}\tau_1 \, a^{-1}(\tau_1)G_k(\tau, \tau_1) \Big|_{-k\tau \to 0} \int \mathrm{d}\tau_2 \, a^{-1}(\tau_2)G_k(\tau, \tau_2) \Big|_{-k\tau \to 0}\nonumber \\
	&\quad  \cdot \int \mathrm{d}^3p \, p_+^2 p_-^2 \tau_1 \tau_2  \left[ f(E_p, \mu) + f(E_p, -\mu) \right]\left[ f(E_{\mathbf{k}-\mathbf{p}}, \mu) + f(E_{\mathbf{k}-\mathbf{p}}, -\mu) \right]  \nonumber \\
	&\quad  \cdot \left| \mathrm J_{\frac12-\mathrm{i}\tilde{m}}(-p\tau_1)\mathrm J_{-\frac12+\mathrm{i}\tilde{m}}(-p\tau_2)
        -\mathrm J_{-\frac12-\mathrm{i}\tilde{m}}(-p\tau_1)\mathrm J_{\frac12+\mathrm{i}\tilde{m}}(-p\tau_2) \right|^2. 
\end{align}
Define the variables $z_1=-k\tau_1$, $x=p/k$, and $z_2=-k\tau_2$, we finally have the correlation function 
\begin{align}
    &\left\langle \hat{h}_{\lambda}^{(s)}(\boldsymbol{k}, 0) \hat{h}_{\lambda}^{(s)}(\boldsymbol{k}', 0)^\dagger \right\rangle \nonumber\\  
    &=\frac{H^4}{M_P^4} \frac{\pi^3}{\cosh^2\pi \tilde{m}} k^{-3} \delta^3(\boldsymbol{k} - \boldsymbol{k}') 
	   \int_0^\infty \mathrm{d}x \int_0^\infty \mathrm{d}z_1 \int_0^\infty \mathrm{d}z_2 \, x^4 z_1 z_2 (\sin z_1 - z_1 \cos z_1) ( \sin z_2 - z_2 \cos z_2 ) \nonumber\\
	&\quad  \cdot \left| \mathrm J_{\frac12-\mathrm{i}\tilde{ m}}(xz_1)\mathrm J_{-\frac12+\mathrm{i}\tilde{ m}}(xz_2)
        -\mathrm J_{-\frac12-\mathrm{i}\tilde{ m}}(xz_1)\mathrm J_{\frac12+\mathrm{i}\tilde{ m}}(xz_2) \right|^2   \nonumber\\
    &\quad  \cdot \left[\frac{1}{\mathrm e^{\left( \sqrt{x^2 z_1^2 + \tilde{m}^2} + \tilde{m} \right) H/T}+1} + \frac{1}{\mathrm e^{\left( \sqrt{x^2 z_1^2 + \tilde{m}^2} - \tilde{m} \right) H/T}+1}\right]  \nonumber \\
	&\quad  \cdot \int_0^\pi\mathrm{d}\theta\sin^5\theta \left[ \frac{1}{\mathrm e^{\left(\sqrt{z_2^2(1-2x\cos\theta+x^2) + \tilde{m}^2}+ \tilde{m} \right) H/T} + 1}
		+ \frac{1}{\mathrm e^{\left(\sqrt{z_2^2(1-2x\cos\theta+x^2) + \tilde{m}^2} - \tilde{m} \right) H/T} + 1} \right], \label{t_correl} 
\end{align}
where we used $\mu=m_{\text{eff}}=m_\psi+g\phi$ for the ideal relativistic fluid and the polarized momentum $\hat p_+\hat p_-=\hat p_x^2+\hat p_y^2=\sin^2\theta$.

\acknowledgments

This work was supported by the National Natural Science Foundation of China (Grant No. 12275143), Central Guidance for Local Science and Technology Development Fund Project (Grands No. 2024ZY0113, 2025ZY0020),
Inner Mongolia Natural Science Foundation (Grants No. 2024SHZR0009, 2026LHMS0083, 2026LHQC0206).



\end{document}